\documentclass[aps,twocolumn,showpacs,pra,amsmath,amssymb,floatfix,superscriptaddress]{revtex4-2}

\usepackage[colorlinks=true,citecolor=blue,linkcolor=magenta,urlcolor=blue]{hyperref}
\usepackage{amsmath,amssymb,graphicx,color}
\usepackage{amsmath}
\usepackage{mathrsfs}

\usepackage{times}
\usepackage[table]{xcolor}
\usepackage{color,cancel}
\usepackage{float}
\usepackage[caption=false]{subfig}
\usepackage{empheq}

\newcommand\ket[1]{\left|\textstyle{#1}\right\rangle}
\newcommand\bra[1]{\left\langle\textstyle{#1}\right|}
\newcommand\braket[1]{\left\langle\textstyle{#1}\right\rangle}

\newcommand\lr{\lambda_r}
\newcommand\lcr{\lambda_{cr}}
\newcommand\gr{g_r}
\newcommand\gcr{g_{cr}}
\newcommand\adag{a^\dagger}

\begin{document}
\title{Multicritical dissipative phase transitions in the anisotropic open quantum Rabi model}

\author{Guitao Lyu}
\affiliation{Division of Natural and Applied Sciences, Duke Kunshan University, Kunshan, Jiangsu 215300, China}

\author{Korbinian Kottmann}
\affiliation{ICFO - Institut de Ciencies Fotoniques, The Barcelona Institute of Science and Technology,Av. Carl Friedrich Gauss 3, 08860 Castelldefels (Barcelona), Spain}

\author{Martin B. Plenio} \email{martin.plenio@uni-ulm.de}
\affiliation{Institute of Theoretical Physics and IQST, Albert-Einstein-Allee 11, Ulm University, D-89081 Ulm, Germany}

\author{Myung-Joong Hwang}	\email{myungjoong.hwang@duke.edu}
\affiliation{Division of Natural and Applied Sciences, Duke Kunshan University, Kunshan, Jiangsu 215300, China}
\affiliation{Zu Chongzhi Center for Mathematics and Computational Science, Duke Kunshan University, Kunshan, Jiangsu 215300, China}

\date{\today}

\begin{abstract}
We investigate the nonequilibrium steady state of the anisotropic open quantum Rabi model, which exhibits first-order and second-order dissipative phase transitions upon varying the degree of anisotropy between the coupling strengths of rotating and counterrotating terms. Using both semiclassical and quantum approaches, we find a rich phase diagram resulting from the interplay between the anisotropy and the dissipation. First, there exists a bistable phase where both the normal and superradiant phases are stable. Second, there are multicritical points where the phase boundaries for the first- and second-order phase transitions meet. We show that a new set of critical exponents governs the scaling of the multicritical points. Finally, we discuss the feasibility of observing the multicritical transitions and bistability using a pair of trapped ions where the anisotropy can be tuned by controlling the intensity of the Raman transitions. Our study enlarges the scope of critical phenomena that may occur in finite-component quantum systems, which could be useful for applications in critical quantum sensing.
\end{abstract}

\maketitle

\section{Introduction}
The investigation of open quantum systems has gained significant popularity due to its fundamental importance and potential applications \cite{Breuer2002Oxford, Muler2012AAMO, RotterRPP2015, Andrew2014AdvPhys, Sieberer2016RPP,Yuto2020AdvPhys}. Phase transitions of the nonequilibrium steady state, known as dissipative phase transitions (DPTs), emerged as a highly important topic within this field \cite{Davide2021RPP, Raghunandan2018PRL, Minganti2021NJP, Casteels2017PRA, Krimer2019PRL, Kessler2012PRA, Drummond1980JPA, Wolinsky1988PRL, Hines2005PRA, Bartolo2016PRA, Minganti2018PRA, Landi2022PRA}. For example, the investigation into the role of criticality on nonequilibrium thermodynamics such as entropy production has gained significant attention~\cite{Tome2012PRL, Shim2016PRE, Zhang2016JSM, Brunelli2018PRL, Herpich2018PRX, Goes2020PRR}.
Recent experimental studies have successfully observed DPTs in a variety of systems, such as semiconductor microcavities~\cite{Rodriguez2017PRL,Fink2018NatPhys,Li2022PRL}, atomic Bose-Einstein condensates (BECs) in optical lattices \cite{ Takafumi2017SciAdv,Benary2022NJP}, waveguide quantum electrodynamics setups \cite{Sheremet2023RMP}, and superconducting circuits \cite{Fitzpatrick2017PRX, Fink2017PRX}. The superradiant phase transition occurring in the Dicke model, observed in systems such as an atomic BEC trapped in an optical cavity~\cite{Baumann2010nature,Baumann2011PRL,Ferdinand2013PNAS,Jens2015PNAS,Baden2014PRL, Hamner2014NatComm, Benjamin2017Natcomm, Kroeze2018PRL, Farokh2021AdvPhys}, is an important example of DPTs. By introducing various types of coherent interactions among a few cavity modes and several BECs, superradiant phase transitions with a wide range of critical phenomena have been predicted and experimentally observed. Among them are multicritical phenomena induced by dissipation and anisotropy~\cite{ETH2018PRL, ETH2021PRX, Keeling2010PRL, Bhaseen2012PRA, Zhang2017Optica,Morales2019PRA,Stitely2020PRA, Baksic2014PRL, Fan2014PRA, Leonard2017Nature}, PT symmetry breaking phase transition due to the nonreciprocal interaction~\cite{Nunnenkamp2023PRL,Nunnenkamp2019PRL}, the emergence of supersolid and spin-glass phase of the BECs~\cite{Benjamin2021PRX,Benjamin2023arXiv}, and frustrated superradiant phase transitions in a Dicke lattice model~\cite{Jinchen2022PRL,Jinchen2023PRR}.

Phase transitions occurring in a system with a finite number of components, far from the traditional thermodynamic limit of infinite particles, have been recently discovered, which is dubbed as finite-component system phase transition~\cite{Hwang2015PRL,Hwang2016PRL}. A prominent example is the quantum Rabi model (QRM) where a single spin is coupled to a single harmonic oscillator~\cite{Hwang2015PRL,Ashhab2013PRA,Bakemeier2012PRA, Cai2021NatComm}, in which the infinite ratio of the qubit transition frequency $\Omega$ and the oscillator frequency $\omega_0$, i.e. $\Omega/\omega_0 \rightarrow \infty$, plays the role of thermodynamic limit. The subsequent research~\cite{Hwang2018PRA} has shown that the open version of the QRM exhibits a DPT, demonstrating the potential of the open QRM as  a promising framework for studying dissipative quantum phase transitions. This approach is particularly promising, as it enables the realization of DPTs in small-scale controlled quantum systems, such as ion traps with only a few ions, where a wide range of coherent interactions, external drivings, and dissipative processes can be engineered~\cite{Cai2022CPL,Schneider2012RPP, Blatt2012NatPhys, Lv2018PRX, Behrle2023PRL}. The interplay among these factors may, in turn, give rise to a variety of nonequilibrium critical phenomena associated with DPTs. Moreover, it has been recently shown that the finite-component system phase transitions occurring in both closed and open systems could be a useful resource for critical quantum sensing~\cite{Garbe2020PRL,Chu2021PRL,Ilias2022PRXQuantu, Ilias2023arXiv}. Therefore, it is an important task to discover and realize DPTs with various phenomenology and universality classes, which could be utilized for developing critical sensing protocols.

Motivated by these opportunities, in this paper we investigate a generalized version of the open QRM where the anisotropy between coupling strength between the rotating and counterrotating terms are considered. The anisotropic open QRM features two fundamentally different coherent processes. The rotating term preserves the total number of excitations of the qubit and the oscillator, while the counterrotating term does not preserves them. Therefore, the nature of nonequilibrium steady state strongly depends on the degree of anisotropy, when the oscillator is coupled to a Markovian bath that constantly removes excitations from the system to the bath. We first find that there are critical values for the anisotropy, beyond and below which no DPTs occur. For the intermediate values of the anisotropy, we find a critical coupling strength at which a second-order DPT occurs that belongs to the same universality class with the isotropic open QRM. Strikingly, upon further increasing the coupling strength beyond the critical coupling, there occurs another DPT of the first-order where the normal phase (NP) reemerges deep in the superradiant phase (SP). This leads to a bistable phase where both the NP and the SP coexist.

The phase boundaries for the second-order and first-order DPTs meet in two points, giving rise to multicritical points. We find that the critical exponents governing these multicritical points are different from those of the second-order DPTs. We note that a similar phase diagram, including the bistable phase and multicritical points, was predicted and observed in the atomic BEC in the cavity system~\cite{ETH2018PRL,ETH2021PRX, Keeling2010PRL}; however, the critical scaling of the tricritical point in this system has not yet been investigated. Here we calculate the critical exponents of the observed tricriticality in the cavity-BEC system and find that the critical exponents are identical with ours, showing that they belong to the same universality class. Finally, we discuss how the multicritical DPT and bistability predicted in the anisotropic open QRM could be realized using two trapped ions.

The paper is organized as follows. In Sec.~\ref{sec:model}, we introduce the anisotropic open QRM. In Sec.~\ref{sec:Semiclassical Analysis}, we provide a semiclassical analysis under the mean-field approximation. The phase diagram is obtained after a stability analysis of the mean-field solutions, and the nature of the DPTs is discussed in Sec.~\ref{sec: Phase Diagram}. In Sec.~\ref{sec: Quantum Solutions}, we provide a full quantum solution for the NP and the SP, respectively. In Sec.~\ref{sec: Universality}, the critical scaling of the vanishing asymptotic decay rate and the diverging oscillator excitation number are investigated. In Sec.~\ref{sec: Implementation}, we propose an experimental scheme for implementing our model using a trapped ion pair. Finally, we draw a conclusion in Sec.~\ref{sec: Conclusion}.

\section{Model} \label{sec:model}
We consider an anisotropic open quantum Rabi model where the rotating and counterrotating terms have different coupling strength, whose dynamics is governed by a master equation,
\begin{equation}
	\label{eq:d_rho}
	\dot \rho=-i[H,\rho]+\kappa D[a]\rho.
\end{equation}
The coherent dynamics of the system is determined by the Hamiltonian, which reads ($\hbar=1$)
\begin{equation}
	\resizebox{1\linewidth}{!}{$H=\omega_0a^\dagger a+\frac{\Omega}{2}\sigma_z-\lambda_r(a\sigma_++a^\dagger\sigma_-)-\lambda_{cr}(a\sigma_-+a^\dagger\sigma_+),$}
	\label{eq:Hamiltonian}
\end{equation}
where $a(a^\dagger)$ is the annihilation (creation) operator of the harmonic oscillator (e.g., a cavity-photon field or a vibrational mode of a trapped ion), and $ \sigma_z $ and $\sigma_\pm \equiv (\sigma_x \pm i\sigma_y)/2$ are Pauli matrices for the two-level system (qubit or spin). The oscillator frequency is $\omega_0$ and the transition frequency for the qubit is $\Omega$.  The coupling strengths are denoted by $\lambda_{c}$  and $\lambda_{cr}$ for the rotating and counterrotating terms, respectively. Following the approach taken in Ref.~\cite{Hwang2018PRA} for the open QRM, we treat the environment for the harmonic oscillator as a local Markovian bath; therefore, the dissipator of the system takes a Lindblad form $D[a]\rho \equiv 2a\rho a^\dagger - a^\dagger a \rho - \rho a^\dagger a$ with a damping rate $\kappa$. Despite the strong qubit-oscillator coupling, such a local dissipator has been shown to be a correction description in the large frequency-ratio limit ~\cite{Ilias2022PRXQuantu}.

The Hamiltonian in Eq.~(\ref{eq:Hamiltonian}) possesses different symmetries associated with the conservation of the total number of excitation, $N_\textrm{tot}=a^\dagger a +(\sigma_z+1)/2$, depending on the relative strength of the coupling terms. For $\lambda_{cr}=0$, it becomes the Jaynes-Cummings (JC) Hamiltonian where $N_\text{tot}$ is the conserved quantity giving rise to the $U(1)$ symmetry. Note that $\lambda_{r}=0$ leads to an anti-JC Hamiltonian, which also possesses the $U(1)$ symmetry. For $\lambda_{cr},\lambda_{r}\neq0$, the parity of the total number of excitation $P=e^{i\pi N_\textrm{tot}}$ is the conserved quantity, leading to the $Z_2$ symmetry. For both cases, the model exhibits a quantum phase transition in the limit $\Omega/\omega_0 \rightarrow \infty$~\cite{Hwang2015PRL,Hwang2016PRL,LiuPRL2017}. For $\lambda_{cr}=\lambda_{r}$, the QRM undergoes a $Z_2$ symmetry-breaking superradiant phase transition~\cite{Hwang2015PRL}. For $\lambda_{cr}(\lambda_{r})=0$, on the other hand, the JC (anti-JC) Hamiltonian undergoes a $U(1)$ symmetry-breaking phase transition where the Goldstone mode emerges as an elementary excitation~\cite{Hwang2016PRL}. Between these two limits where the nonzero coupling strength $\lambda_{cr}$ and $\lambda_{r}$ are not equal, the generalized QRM still possesses the $Z_2$ symmetry and therefore the nature of ground-state phase transition including their critical exponents does not change from that of the symmetric case ($\lambda_{cr}=\lambda_{r}$)~\cite{LiuPRL2017}, as one would expect from the perspective of universality. The phase boundary of the ground-state phase transition is simply determined by $\lambda_r+\lambda_{cr}=\sqrt{\omega\Omega}$.

In the presence of dissipation, Ref.~\cite{Hwang2018PRA} showed that the steady state undergoes a DPT for the symmetric case ($\lambda_r=\lambda_{cr}$) of Eq.~(\ref{eq:d_rho}). The DPT of the open QRM has a critical point that is shifted by the dissipation, $\frac{\sqrt{\omega_0\Omega}}{2}\sqrt{1+\kappa^2/\omega_0^2}$, and the critical exponent for the diverging excitation number of the oscillator changes from $1/2$ of the closed QRM to $1$ of the open QRM. On the other hand, it is straightforward to anticipate the role of dissipation on the quantum phase transition of the JC model, predicted for the $\kappa=0$ case in Ref.~\cite{Hwang2016PRL}. Namely, the open JC model with $\lambda_{cr}=0$ and $\kappa\neq0$ does not go through a DPT, because the Hamiltonian has only particle number conserving interactions while the dissipation keeps removing excitations from the system until it becomes a simple vacuum state. The fundamentally different role of the dissipation on the steady-state diagram in these two limits ($\lambda_{cr}=\lambda_{r}$ and $\lambda_{cr}=0$) suggests that the competition between the dissipation and the counterrotating term may give rise to a phase diagram that is strikingly different from the ground-state phase diagram. Therefore, understanding the steady-state phase diagram and their critical properties for the anisotropic open QRM for varying $\lambda_{cr}$ and $\lambda_{r}$ is the goal of the present paper. We note that in the experimental realization of the open QRM using ion-trap~\cite{Hwang2018PRA,Cai2022CPL}, it is straightforward to control the relative strength of the rotating and counterrotating terms by modulating the strength of the red and blue sideband transitions, respectively (see the implementation in Sec.~\ref{sec: Implementation}). Moreover, the Lindblad master equation for the open QRM with the local dissipator can be derived from the microscopic models.

\section{Semiclassical Analysis} \label{sec:Semiclassical Analysis}
In this section, we find a mean-field solution for the steady states by solving the semiclassical equation of motion, which exactly captures the phase diagram of the system in the limit $\Omega/\omega_0 \rightarrow \infty$. The nature of quantum fluctuations around the mean-field solution will be discussed in the following sections. By neglecting the quantum fluctuations and factorizing the expectation values of the operators, the mean-field equations of motion following Eq.~(\ref{eq:d_rho}) are given as follows:
\begin{align}
	\label{eom1}\dot{\braket{a}}= & -i (\omega_0-i\kappa) \braket{a}+i(\lcr \braket{\sigma_+}+\lr \braket{\sigma_-}),\\
	\label{eom2}\dot{\braket{\sigma_+}}=&i \Omega\braket{\sigma_+}+i(\lcr \braket{a}+\lr \braket{a^\dagger})\braket{\sigma_z},\\
	\label{eom3}\dot{\braket{\sigma_z}}=&i 2(\lr \braket{a}+\lcr\braket{a^\dagger})\braket{\sigma_+} + \textrm{c.c.}
	%\nonumber\\&-i2(\lcr \braket{a}+\lr\braket{a^\dagger})\braket{\sigma_-},
\end{align}
where, for example, $\dot{\braket{a}} \equiv \frac{d}{dt} \braket{a}$ and $\braket{a} \equiv \mathrm{Tr}[\rho a]$.

It is convenient to define several renormalized parameters for our analysis. We define the dimensionless coupling strength $\gr \equiv \lr / \sqrt{\omega_0\Omega}$ and $\gcr \equiv \lcr / \sqrt{\omega_0\Omega}$ and the dimensionless decay rate $\bar{\kappa} \equiv \kappa/\omega_0$. We denote the mean value of the oscillator coherence as $\alpha\equiv x+iy \equiv \braket{a}$, and the renormalized one as $\bar{\alpha}\equiv \bar{x}+i\bar{y} \equiv \sqrt{\frac{\omega_0}{\Omega}}\braket{a}$.
Finally, for spin expectation values, $s_{x,y,z}\equiv\braket{\sigma_{x,y,z}}$ and $s_+=(s_x+is_y)/2$.
Then, the steady-state solution satisfies
\begin{align}
	\label{0eom1}0&=-(1-i\bar\kappa) \bar{\alpha}+ \gr s_+^*+ \gcr s_+,\\
	\label{0eom2}0&= s_++ (\gr \bar{\alpha}^*+\gcr \bar{\alpha})s_z,\\
	\label{0eom3}0&=(\gr \bar{\alpha}+\gcr \bar{\alpha}^*)s_+-(\gr \bar{\alpha}^*+ \gcr \bar{\alpha})s_+^*.
\end{align}

Note that Eq.~(\ref{0eom3}) is trivially fulfilled when Eq.~(\ref{0eom2}) is satisfied. Thus, from Eqs.~(\ref{0eom1}) and (\ref{0eom2}), we obtain a system of four linear equations, for the variables $\bar{x}, \bar{y}, s_x$, and $s_y$, parametrized by $s_z$. That is,
\begin{equation}
	\label{eq:Lcl_0}
	\renewcommand{\arraystretch}{1.1}
	L_{cl}\cdot \left(
		\bar{x},~\bar{y},~s_x,~s_y
	\right)^\intercal=0,
\end{equation}
with
\begin{align}
	\label{eq:Lcl}
	\renewcommand{\arraystretch}{1.2}
	\resizebox{1\linewidth}{!}{$
		L_{cl}=\left(\begin{array}{cccc}
			1 & \bar{\kappa} & -g(1+\varepsilon)/2 & 0 \\
			\bar{\kappa} & -1 & 0 & -g(1-\varepsilon)/2 \\
			g(1+\varepsilon) s_z & 0 & 1/2 & 0 \\
			0 & -g(1-\varepsilon) s_z & 0 & 1/2
		\end{array}\right),$}
\end{align}
where we have used the definition
\begin{align}
g_r \equiv g, \ g_{cr} \equiv \varepsilon g.
\label{eq:g}
\end{align}
The parameter $\varepsilon$  represents the asymmetry in the coupling strength between the rotating and counterrotating terms. Our model returns to the Rabi model as $\varepsilon=1$, and goes to the JC model as $\varepsilon=0$.

There is a trivial solution for Eq.~(\ref{eq:Lcl_0}),
\begin{equation}
	\label{eq:np_solution}
	\bar{x} = \bar{y} = s_x = s_y = 0, s_z=\pm1.
\end{equation}
In the following, we only consider the $s_z=-1$ solution, which we refer to as the NP solution, since $s_z=+1$ corresponds to infinitely energetic state in the limit of $\Omega/\omega_0\rightarrow\infty$.

On the other hand, the system has nontrivial solutions only when the determinant of  Eq.~(\ref{eq:Lcl}) is zero, i.e.,
\begin{equation}
	\resizebox{1\linewidth}{!}{$\det[ L_{cl}]=-\dfrac{1}{4}[1+\bar{\kappa}^2+2 \left(\varepsilon^2+1\right) g^2 s_z +\left(\varepsilon^2-1\right)^2 g^4s_z^2]= 0.$}
	\label{eq:det_Lcl_0}
\end{equation}
This leads to a nontrivial solution for the $s_z$,
\begin{equation}
	\label{eq: s_zsp}
	s_{z} = -\frac{(1+\varepsilon^2)-\sqrt{4\varepsilon^2-\bar{\kappa}^2(1-\varepsilon^2)^2}}{(1-\varepsilon^2)^2g^2}.
\end{equation}
We note that there is another solution of Eq.~(\ref{eq:det_Lcl_0}) which we denote as $s_z^+ \equiv -[(1+\varepsilon^2)+\sqrt{4\varepsilon^2-\bar{\kappa}^2(1-\varepsilon^2)^2}]/(1-\varepsilon^2)^2g^2$ for $\varepsilon\neq1$ , but it is not a stable solution as we show in Appendix~\ref{App:SP with s_z^+}; therefore, we neglect it here. Using the condition of spin conservation $ s_z^2+s_x^2+s_y^2=1$, we find corresponding nontrivial solutions for the renormalized oscillator coherence $\bar{\alpha}=\bar{x}+i\bar{y}$ where
\begin{equation}
	\label{result_x}
	\bar{x} = \pm \sqrt{\frac{1 - s_z^2}{4g^2 s_z^2 \left[ (1+\varepsilon)^2 + \frac{\bar{\kappa}^2 (1-\varepsilon)^2}{[1 + g^2(1-\varepsilon)^2 s_z]^2} \right]}},
\end{equation}
and
\begin{equation}
	\label{result_y}
	\bar{y} = \text{sign}(\bar{x}) \sqrt{\frac{1 - s_z^2}{4g^2 s_z^2 \left[(1-\varepsilon)^2 + \frac{\bar{\kappa}^2 (1 + \varepsilon)^2}{\left[ 1 + g^2 (1+\varepsilon)^2 s_z \right]^2} \right]}}.
\end{equation}
Therefore, we identify the nontrivial solution as the SP where the $Z_2$ symmetry of the system is spontaneously broken. In SP, the spin also acquires a spontaneous polarization, which is given by
\begin{equation}
	\label{eq:s_x_y}
	s_x = -2g(1+\varepsilon) \bar{x} s_z  \   \text{and} \   s_y = 2g(1-\varepsilon) \bar{y} s_z.
\end{equation}

Since $s_z$ must be a real number, there exists a range of asymmetry parameter $\varepsilon$,
\begin{align}
	\varepsilon_{\textrm{min}}(\bar\kappa)\leq\varepsilon\leq\varepsilon_{\textrm{max}}(\bar\kappa),
\end{align}
with
\begin{align}
	\varepsilon_{\textrm{max, min}}(\bar\kappa)=\frac{\pm1+\sqrt{1+\bar\kappa^2}}{\bar\kappa},
	\label{eq:epsilon_max}
\end{align}
within which the SP solution exists. Furthermore, the critical value of the coupling strength $g$ at the phase boundary between the NP and the SP is determined by setting $s_z=-1$ in Eq.~(\ref{eq:det_Lcl_0}), which reads
\begin{equation}
	\label{eq:g_c}
	g_{c}^\pm (\varepsilon, \bar{\kappa})= \sqrt{\frac{(1+\varepsilon^2) \pm \sqrt{4\varepsilon^2-\bar{\kappa}^2(1-\varepsilon^2)^2}}{(1-\varepsilon^2)^2}}
\end{equation}
It is interesting to note that,  for a given asymmetry $\varepsilon$ within $\varepsilon_{\textrm{min}}(\bar\kappa)\leq\varepsilon\leq\varepsilon_{\textrm{max}}(\bar\kappa)$, there are two critical points $g_c^\pm(\varepsilon,\bar{\kappa})$. We will demonstrate below that $g_c^-(\varepsilon,\bar{\kappa})$ corresponds to a second-order dissipative phase transition from the NP to the SP. Strikingly, $g_c^+(\varepsilon,\bar{\kappa})$ corresponds to a point within the SP, beyond which the NP become stable again, leading to a bistable phase (see Fig.~\ref{fig:phasediagram}).

It is instructive to check that our SP solution recovers the results obtained in the symmetric open QRM~\cite{Hwang2018PRA}. That is, $s_z = - \frac{1 + \bar{\kappa}^2}{(2g)^2}$ and $\bar{\alpha} = \pm \frac{g}{1 -i\bar{\kappa}} \sqrt{1 - \frac{(1 + \bar{\kappa}^2)^2}{(2g)^4}}$ as $\varepsilon \rightarrow 1$, where the critical coupling strength $g=g_c^-=\sqrt{1 + \bar{\kappa}^2}/2$. On the other hand, $g_c^+$ diverges at $\varepsilon=1$; it indicates that the bistability is completely absent for the symmetric open QRM and that the physics we discuss below is a unique feature of the anisotropic open QRM.

\section{Bistable phase and Tricritical point} \label{sec: Phase Diagram}
\begin{figure}[tpb!]
	\includegraphics[width=0.49\linewidth]{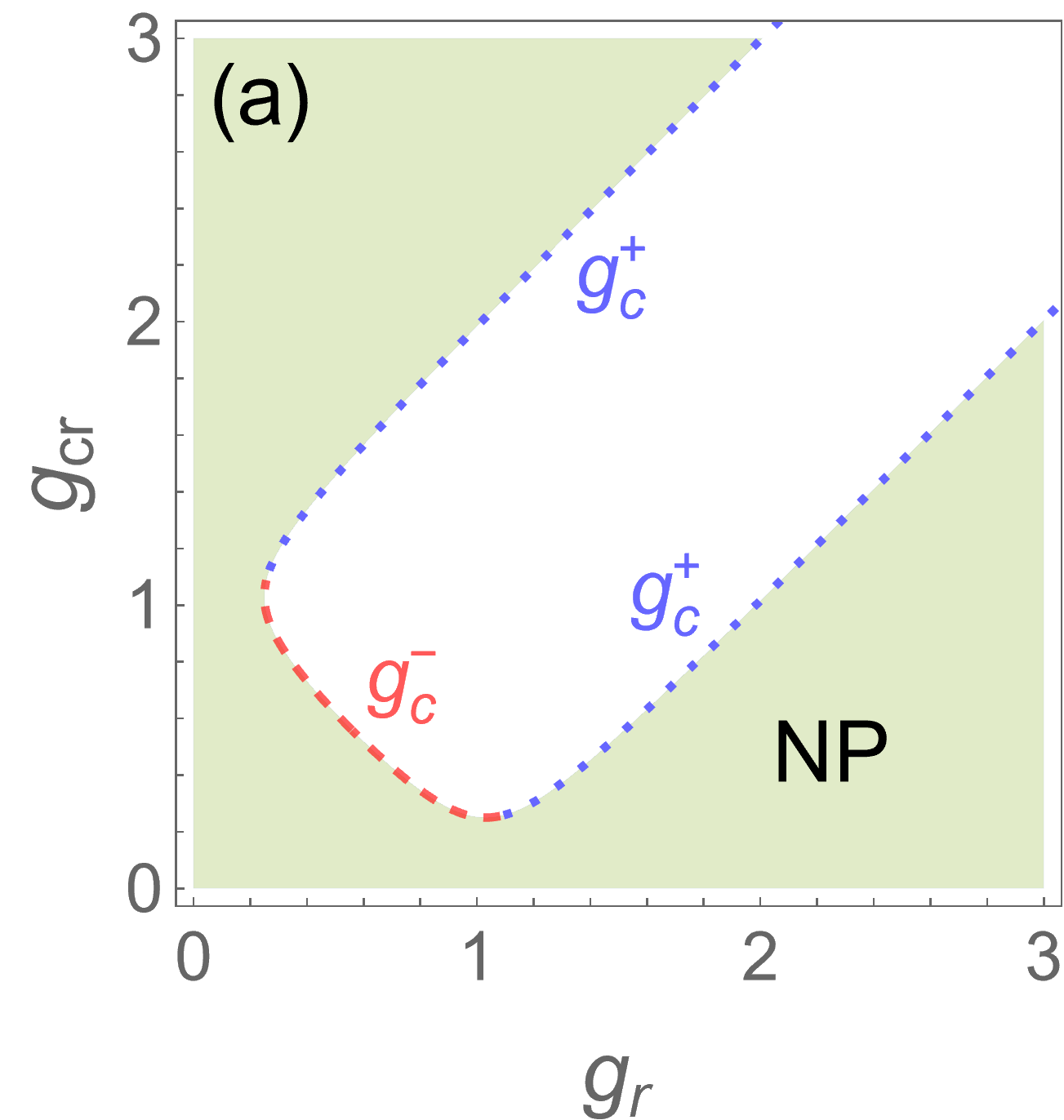}
	\includegraphics[width=0.49\linewidth]{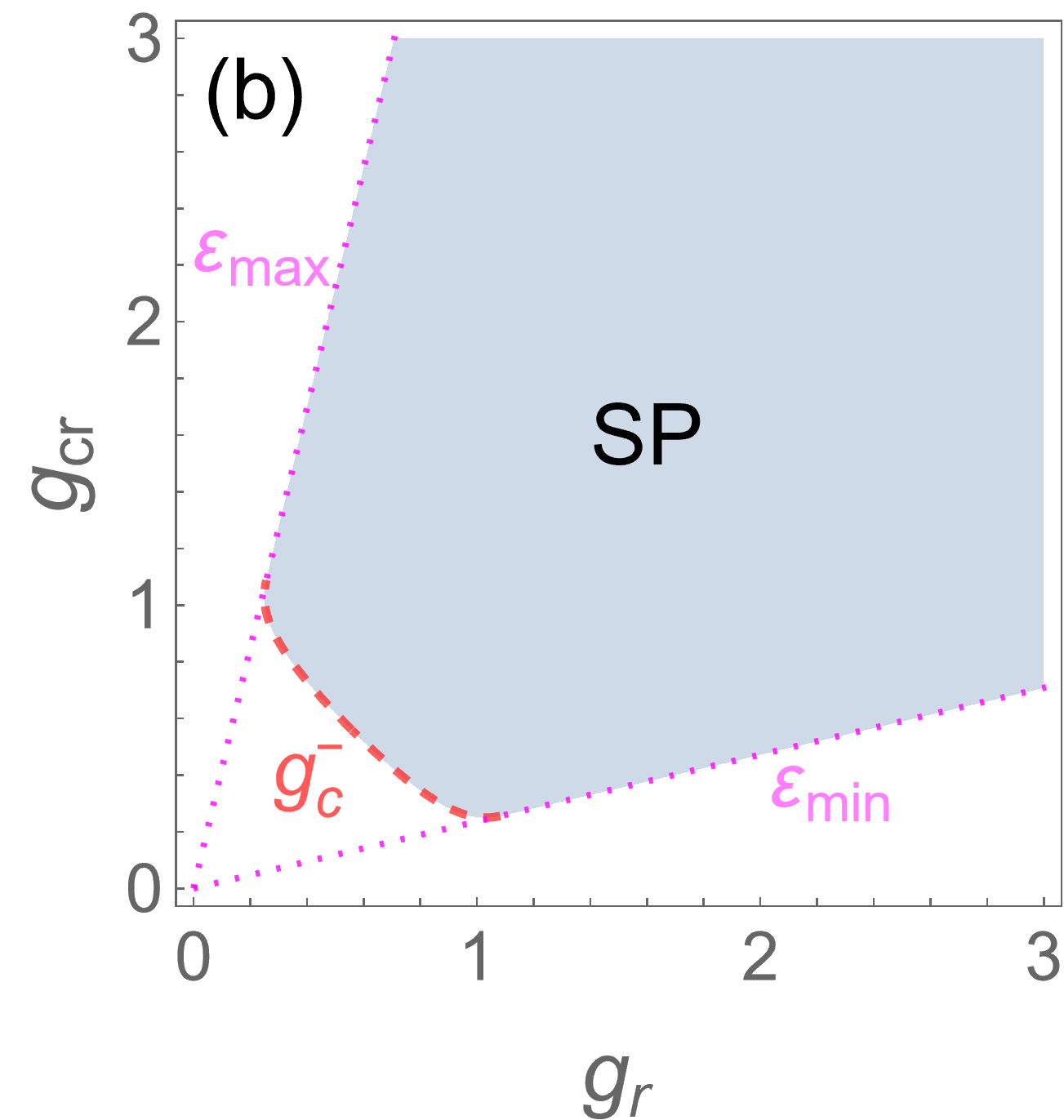}
	\vspace{0.35cm}

	\includegraphics[width=1\linewidth]{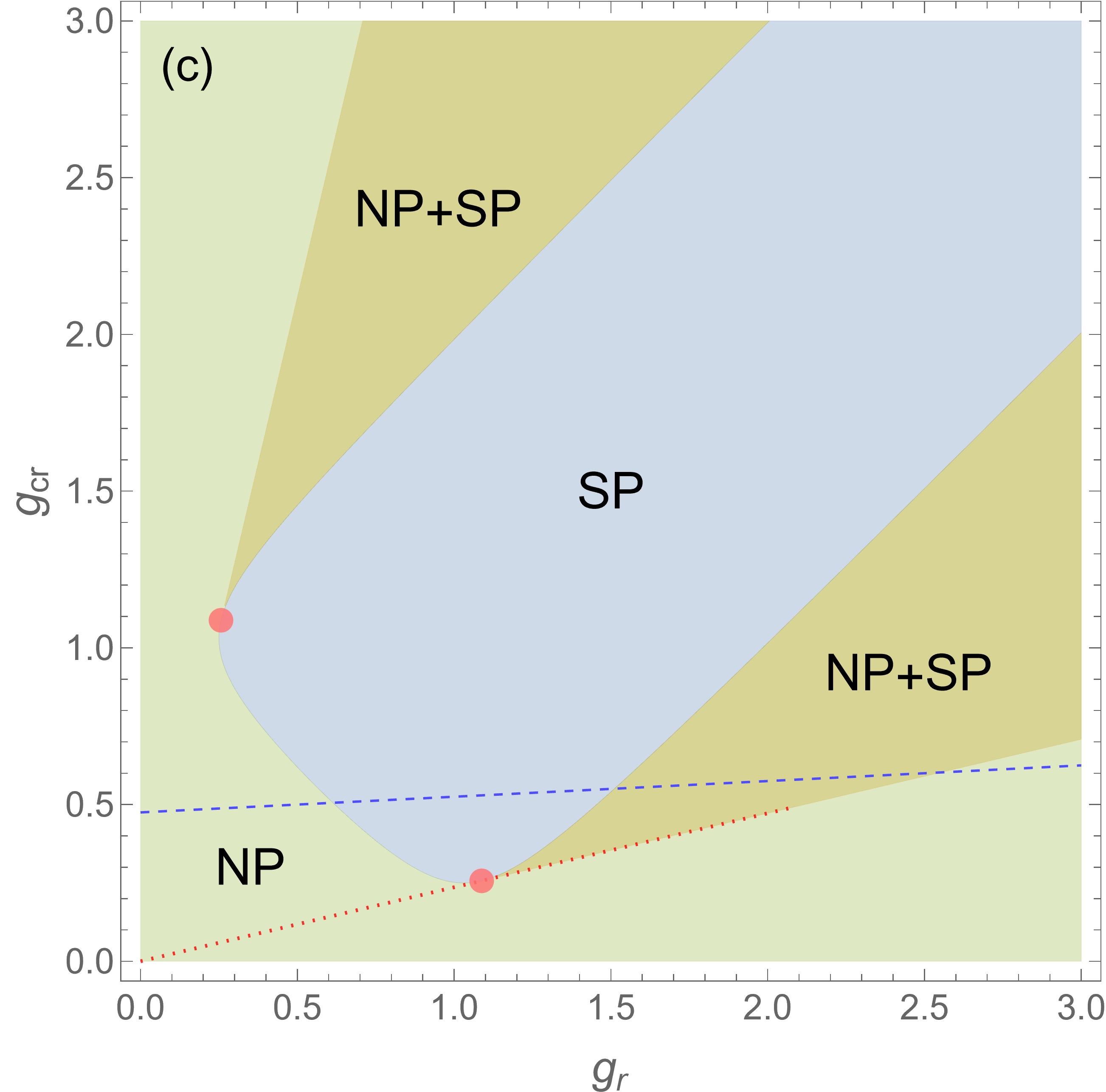}
	\caption{Phase diagram of the anisotropic open quantum Rabi model.
		(a) The stable region for the normal
		phase (NP) (shaded in green). The blue dotted and red dashed curves depict the critical coupling strength $g_c^\pm (\varepsilon, \bar{\kappa})$, respectively.
		(b) The stable region (shaded in blue) for the superradiant phase (SP). The red dashed curve $g_c^- (\varepsilon, \bar{\kappa})$ is the same one in (a). The magenta dotted line has the slope $\varepsilon_\text{min (max)}(\bar{\kappa})$.
		(c) The phase diagram. The overlap region between the NP in (a) and the SP in (b) is a bistable phase shaded in yellow. Two red dots, located at the intersection of three phases, represent the tricritical points. The dashed line $g_{cr}=0.05(g_r-0.5)+0.5$ and dotted line $g_{cr}=g_r\varepsilon_\text{min}$ in (c) are used in later discussions. Here we set $\bar\kappa=0.5$.}
	\label{fig:phasediagram}
\end{figure}

In order to determine the phase diagram for the steady state, we examine the stability of the NP and SP solutions, respectively. We find that the stability condition for the NP reads (see Appendix~\ref{App:Stability Analysis} for details)
\begin{equation}
	\label{eq:A_np}
	A_{\text{NP}}=1 + \bar{\kappa}^2 - 2 (1 + \varepsilon^2) g^2 + (1 - \varepsilon^2)^2 g^4>0.
\end{equation}
From the above condition, we obtain the stable region of NP (shaded in green) as shown in Fig.~\ref{fig:phasediagram}(a). For the phase diagram, we take the decay rate $\bar{\kappa}=0.5$. Note that the condition for the NP boundary, $A_{\text{NP}}=0$, is the same as Eq.~(\ref{eq:det_Lcl_0}) for $s_z=-1$, consequently giving rise to the same expression for the critical coupling strength $g_c^\pm(\varepsilon, \bar{\kappa})$ given in Eq.~(\ref{eq:g_c}) and depicted in Fig.~\ref{fig:phasediagram}(a) by the blue dotted and red dashed curves, which act as the boundaries for the NP. The stability of the NP exhibits striking features. First, for $\varepsilon<\varepsilon_{\textrm{min}}$ and $\varepsilon>\varepsilon_{\textrm{max}}$, the NP is always stable. The line for $\varepsilon_{\textrm{min(max)}}$ is shown in Fig.~\ref{fig:phasediagram}(b). This shows that when the coupling strength of the counterrotating term $g_{cr}$ is much smaller than that of the rotating term, or vice versa, the $U(1)$ character of the steady state persists; namely, the oscillator damping dominates over the particle-number non-preserving interactions and the steady state remains to be vacuum. Second, for $\varepsilon_{\textrm{min}}<\varepsilon<\varepsilon_{\textrm{max}}$, the NP becomes unstable for $g>g_c^-(\varepsilon,\bar{\kappa})$; strikingly, however, as one keeps increasing $g$ there is another critical point $g_c^+(\varepsilon,\bar{\kappa})$ beyond which the NP becomes stable again. The recurrence of the NP leads to a highly non-monotonic phase boundary in the $g_c-g_{cr}$ phase diagram. The SP is stable when $g>g_c^-(\varepsilon,\bar{\kappa})$ for $\varepsilon_{\textrm{min}}<\varepsilon<\varepsilon_{\textrm{max}}$. The stable region (blue shade) for the SP is depicted in Fig.~\ref{fig:phasediagram}(b). $g_c^-(\varepsilon,\bar{\kappa})$ is displayed by a red dashed curve in the panel. The dotted line $g_{cr}=g_r\varepsilon_\text{min (max)}$, with the slope $\varepsilon_\text{min (max)}$, defines the phase boundary for the SP. In addition, it is also worth noting that the trivial solution with $s_z=+1$ in Eq.~(\ref{eq:np_solution}) is found to be stable for any values of $g$, $\varepsilon$, and $\bar{\kappa}$ by examining the coefficient $A$ in Appendix~\ref{App:Stability Analysis}, while it is not shown in the phase diagram.

%---------------------------------
%---------------------------------

The resulting rich phase diagram of the anisotropic open QRM is shown in Fig.~\ref{fig:phasediagram}(c). There are two notable features. First, a bistable phase emerges where the NP and SP phases coexist (shaded in yellow). In the symmetric case ($g_{cr}=g_r$), the system exhibits a transition from the NP to the SP when the coupling strength reaches the critical value ($g_c^-=\sqrt{1 + \bar{\kappa}^2}/2$) and the NP remains unstable for any value of $g>g_c^-$~\cite{Hwang2018PRA}. However, by introducing an asymmetry in the coupling terms, the NP becomes stable again as one increases the coupling strengths even further beyond the critical value $g_c^+(\varepsilon, \bar{\kappa})$, i.e. as $g>g_c^+(\varepsilon, \bar{\kappa})$, coexisting with the SP. Second, there is a tricritical point (red dot) located at the intersection of the three phases, where the phase boundary curve $g_c^-(\varepsilon,\bar{\kappa})$ (a boundary of a second-order phase transition) meets the phase boundary line $g_{cr}=g_r\varepsilon_\text{min (max)}$ (a boundary of a first-order phase transition).

\begin{figure}[tpb!]
	\includegraphics[width=0.95\linewidth]{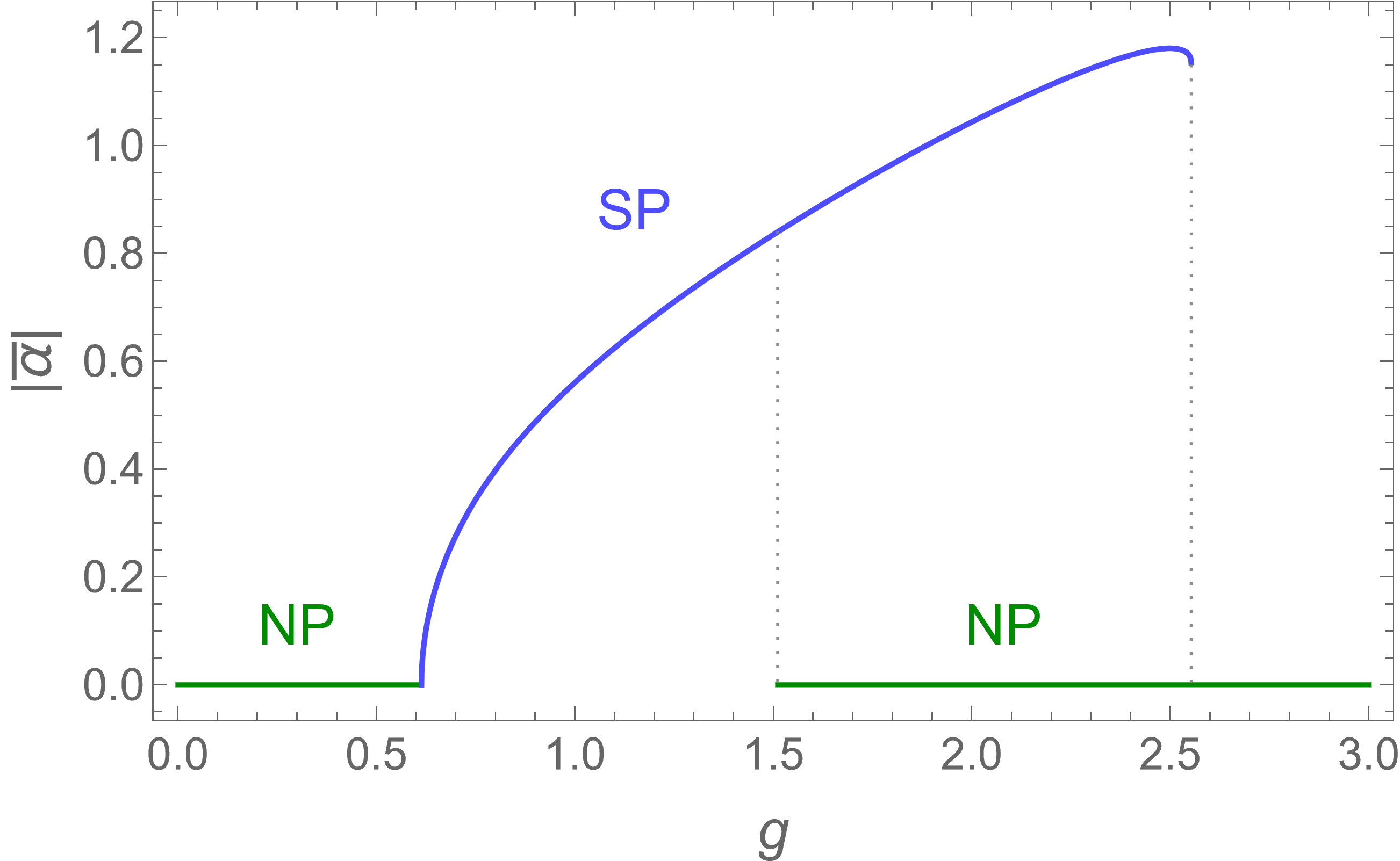}
	\caption{First- and second-order DPT and hysteresis due to bistability. The absolute value of the order parameter $\bar{\alpha}$ as a function of $g$  (namely $g_r$) is depicted as one moves along the blue dashed line in Fig.~\ref{fig:phasediagram}(c).
	}
	\label{fig:a_g}
\end{figure}

In order to demonstrate the first-order and second-order DPT of the system, we plot in Fig.~\ref{fig:a_g} the renormalized order parameter $\bar{\alpha}$ as $g$ (namely $g_r$) traverses the normal and superradiant phases along the blue dashed line in Fig.~\ref{fig:phasediagram}(c). When $g$ goes a from the NP to the SP, the absolute value of $\bar{\alpha}$ presents a second-order phase transition at the critical coupling strength $g_c^-\approx 0.61$. For convenience, we denote the value of $g$ at the phase boundary line $g_{cr}=g_r\varepsilon_\text{min (max)}$ as $g_{\varepsilon_\text{min (max)}}$. When crossing the phase boundary line at $g=g_{\varepsilon_\text{min}}\approx 2.55$, the order parameter $\bar{\alpha}$ turns to zero abruptly, leading to a discontinuity of the order parameter and the first-order phase transition on the boundary line. We also note that there are two values of $|\bar{\alpha}|$ in the range of the bistable phase from the critical coupling strength $g_c^+\approx 1.51$ to the phase boundary line at $g_{\varepsilon_\text{min}}\approx 2.55$. Interestingly, due to the presence of the bistable phase, a hysteresis effect could be observed at the boundary between the normal and superradiant phases when $g$ crosses the bistable phase forward and backward~\cite{Stitely2020PRA}. A similar hysteresis between the NP and SP has been experimentally observed using a driven BEC in a cavity system~\cite{ETH2021PRX}.

In Figs.~\ref{fig:phasediagram} and \ref{fig:a_g} we have chosen $\bar\kappa=0.5$. As one decreases $\bar\kappa$, the tricritical points approach the $g_{cr}$ and $g_r$ axes; therefore, the area of the NP shrinks. At the same time, the first-order phase transition boundary, given by $\varepsilon_{\textrm{min (max)}}(\bar\kappa)$, comes closer to the $g_c$ and $g_{cr}$ axis, leading to the widening of the SP. As one increases $\bar\kappa$, the trend is reversed.

Here, let us compare the phase diagram of the anisotropic open QRM in Fig.~\ref{fig:phasediagram} with that of the anisotropic open Dicke model~\cite{ETH2018PRL, Keeling2010PRL, Stitely2020PRR}. The mean-field equations of motion of the latter (e.g., Eq.~(9) in Ref.~\cite{Stitely2020PRR}) can be rescaled to remove the dependence on the number of spins, which then becomes exactly the same equations as Eqs.~(\ref{eom1}-\ref{eom3}). Interestingly, however, the two models exhibit qualitatively different phase diagrams. In the anisotropic open Dicke model, there emerges nonstationary steady states featuring limit-cycles and chaos for $\varepsilon>1$ $(g_{cr}>g_{r})$, a feature that is absent in Fig.~\ref{fig:phasediagram}.
The difference, in fact, comes from the number of dynamical variables. In the open QRM, on the one hand, we consider the limit of $\Omega/\omega_0\rightarrow\infty$, where the spin instantly follows the cavity dynamics and can be adiabatically eliminated. Therefore, the cavity field becomes the only dynamical variable and the steady state is determined by a nonlinear equation of motion of a single oscillator, which cannot exhibit any chaotic behavior. See Appendix~\ref {App:simulation} for more details. In the open Dicke model, on the other hand, the thermodynamic limit is achieved by the infinite number of spins and the relative ratio of frequencies remain finite. In this case, both the spin and the cavity are dynamical variables, whose coupled equations of motion feature nonstationary steady states~\cite{Stitely2020PRR}. It is interesting to note the different roles played by anisotropy for the quantum phase transition and DPT. In closed systems, the quantum phase transition of the QRM and the Dicke model belong to the same universality class for any values of $\varepsilon$~\cite{Hwang2015PRL, LiuPRL2017}. In open systems, however, the DPT of the open QRM and the open Dicke model belongs to the same universality class only when the counter-rotating term is smaller than the rotating terms $(g_{cr}<g_{r})$; for $(g_{cr}>g_{r})$, they exhibit completely different phases.

%------------------------------------
%------------------------------------

\section{Quantum fluctuations}  \label{sec: Quantum Solutions}
Having established the mean-field phase diagram of the anisotropic open QRM, in this section we will provide a full quantum solution for the NP and the SP. First, we derive an effective master equation in the $\Omega/\omega_0 \rightarrow \infty$ limit, characterized by a quadratic form involving the oscillator operator $a$. Utilizing this quadratic effective master equation, we study the critical scaling of quantum fluctuation and asymptotic decay rate near the phase boundaries and the tricritical point. In addition, we complement the quantum fluctuations analysis in the thermodynamic limit presented in this section with numerical simulation of the master equation (\ref{eq:d_rho}) for finite frequency ratio $\eta\equiv\Omega/\omega_0$ and present the results in Appendix~\ref {App:quantum simulation}.

\subsection{Normal phase} \label{subsec: Normal Phase}
By performing the Schrieffer-Wolff (SW) transformation to the Hamiltonian (\ref{eq:Hamiltonian}) and projecting it to the spin-down subspace~\cite{Hwang2015PRL,Hwang2018PRA}, we obtain an effective Hamiltonian for the NP, which reads
\begin{align}
	H_\text{np} &=\omega_0a^\dagger a-\omega_0\left[g_r^2 a^\dagger a +g_{cr}^2 a a^\dagger +g_rg_{cr}(a^{\dagger 2}+a^2)\right].
\end{align}
Note that the higher order terms whose coefficients become zero in the  $\Omega/\omega_0 \rightarrow \infty$ limit are neglected~\cite{Hwang2015PRL,Hwang2018PRA}. Correspondingly, the effective master equation becomes
\begin{equation}
	\dot \rho_\text{np}=-i[H_\text{np},\rho_\text{np}]+\kappa D[a]\rho_\text{np},
\end{equation}
where the density matrix $\rho_\text{np}= \bra{\downarrow} \mathrm{e}^{-S}\rho\mathrm{e}^{S} \ket{\downarrow}$ and the generator for the SW transformation is given by
$S=\frac{\lambda_r}{\Omega}(a\sigma_+-\adag\sigma_-)-\frac{\lambda_{cr}}{\Omega}(a\sigma_--\adag\sigma_+)$. Then, the equation of motion for the first moment of the oscillator operators, $\textbf{a}=(\braket{a},\braket{a^\dagger})^\intercal$, reads
\begin{equation}
	\dot{\textbf{a}}=L_\textrm{np} \textbf{a}
\end{equation}
where
\begin{align}
	\label{Lstability}
	\renewcommand{\arraystretch}{1.5}
	\resizebox{1\linewidth}{!}{$
	L_\textrm{np}=\left(\begin{array}{cc}i \omega_0(g_r^2+g_{cr}^2-1) -\kappa& i2\omega_0 g_rg_{cr} \\-i2\omega_0 g_rg_{cr} & i \omega_0 (1-g_r^2-g_{cr}^2)-\kappa\end{array}\right).
	$}
\end{align}

\begin{figure}[tb!]
	\includegraphics[width=0.975\linewidth]{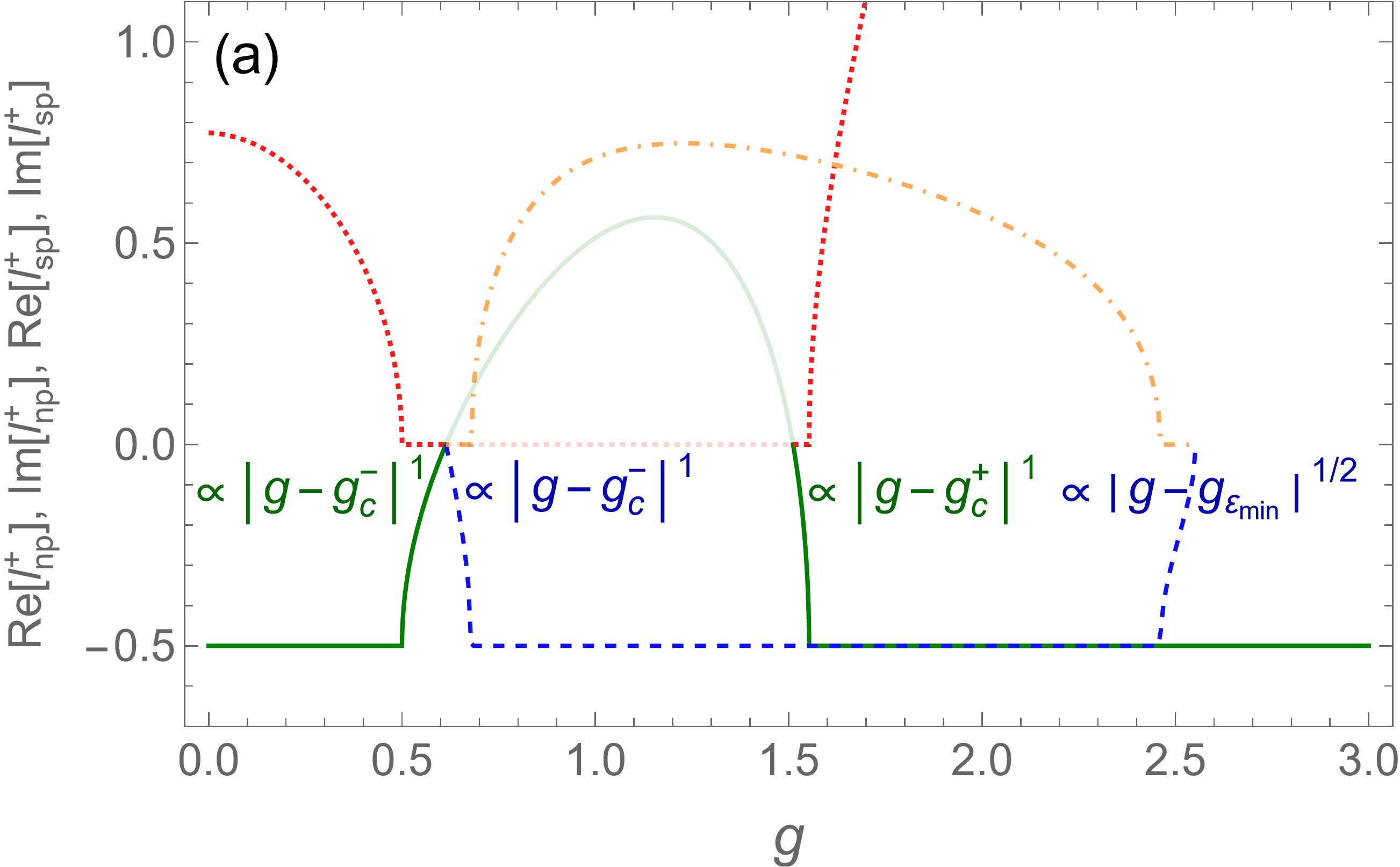}
	\vspace{0.3cm}

	\includegraphics[width=0.975\linewidth]{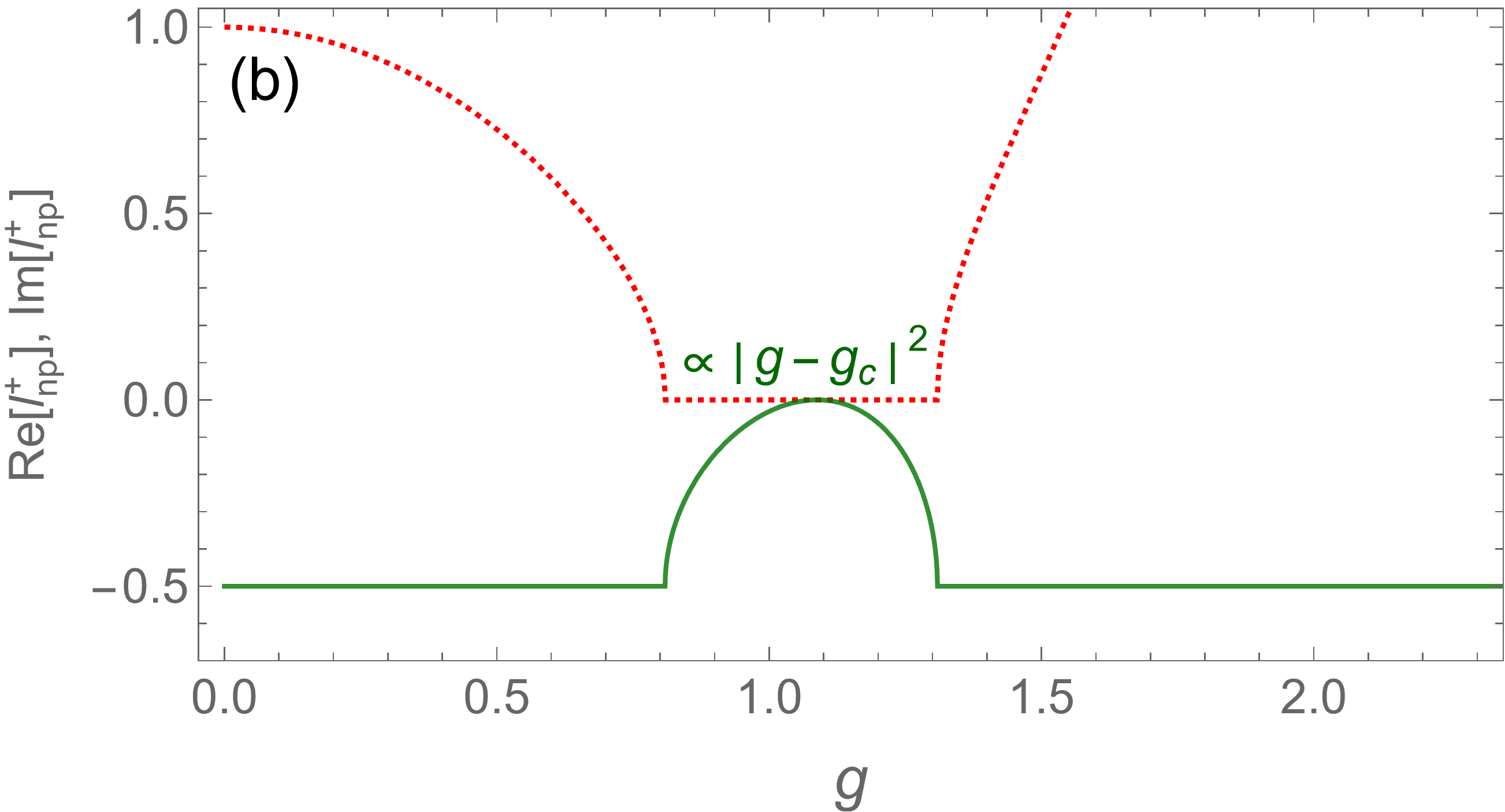}
	\caption{Asymptotic decay rate (ADR) for the NP and the SP. (a) The eigenvalues of dynamical matrices $l^+_\textrm{np (sp)}$ as a function of $g$ are presented as one moves along the blue dashed line in Fig.~\ref{fig:phasediagram}(c). The real (imaginary) part of $l^+_\textrm{np}$ and $l^+_\textrm{sp}$ is given by the green solid (red dotted) curve and blue dashed (orange dash-dotted) curve, respectively. (b) The real and imaginary parts of $l^+_\textrm{np}$ as one moves along the red dotted line in Fig.~\ref{fig:phasediagram}(c), which is the phase boundary line $g_{cr}=g_r\varepsilon_\text{min}$. The scaling of ADR near the critical coupling strengths $g_c^\pm$ and $g_{\varepsilon_\text{min}}$ is presented in each panel. Here $\bar\kappa=0.5$.
	}
	\label{fig:Lnpsp}
\end{figure}

The eigenvalue $l_\textrm{np}$ of $L_\textrm{np}/\omega_0$ reads
\begin{align}
	\label{eq:lnp}
	l_\textrm{np}^\pm=-\bar\kappa\pm  \sqrt{2 g^2 \left(1+\varepsilon^2\right)-g^4 \left(1-\varepsilon^2\right)^2-1}.
\end{align}
The NP is stable as long as the real part of the eigenvalues are negative. This condition agrees with the stable region of NP shown in Fig.~\ref{fig:phasediagram}(a). Since $\mathrm{Re}[l_\textrm{np}^-]$ is always negative, we only need to examine $l_\textrm{np}^+$. In Fig.~\ref{fig:Lnpsp}(a), we plot the real part (green solid curve) and the imaginary part (red dotted curve) of $l_\textrm{np}^+$ for the case $g$ going along the blue dashed line in Fig.~\ref{fig:phasediagram}(c). For small $g$ far away from the critical point ($g \ll g_c^-(\varepsilon,\bar\kappa)$), the square root term is purely imaginary. In this case, the dynamics of the system is underdamped and the decay rate is $\bar\kappa$. As $g$ approaches the critical point, the dynamics becomes overdamped with a zero imaginary part. The effective decay rate in this overdamped region is called as an asymptotic decay rate (ADR),  which tends to zero as $g$ approaches its critical values. This behavior coincides with the closing of the Liouvillian gap near the critical point~\cite{Kessler2012PRA, Fitzpatrick2017PRX,Hwang2018PRA}.

For $g>g_c^-$, $\mathrm{Re}[l_\textrm{np}^+]$ becomes positive, so that the NP becomes unstable. We note, however, that the inner square root part of $l_\textrm{np}^+$ has an inverted parabolic shape with respect to $g^2$ when $\varepsilon\neq1$.Therefore, upon further increasing $g$ with $\varepsilon\neq1$, the real part of $l_\textrm{np}^+$ becomes negative again for $g>g_c^+(\varepsilon,\bar\kappa)$ [see the light solid curve in Fig.~\ref{fig:Lnpsp}(a)]. The recurrence of the NP therefore occurs due to the competition between the bare oscillator decay rate $\bar\kappa$ and the asymmetry $\varepsilon$, which results in a non-monotonic behavior of the ADR in the overdamped dynamics.

We also examine fluctuations of the oscillator excitation in the NP. The dynamics of the second moments $\textbf{s}=\left(\braket{a^\dagger a},\braket{a^2},\braket{a^{\dagger2}}\right)^\intercal$ of the  oscillator is governed by
\begin{align}
	\label{coherence}
	\renewcommand{\arraystretch}{1.2}
	\dot{\textbf{s}}=M_\textrm{np}\textbf{s} +Y_\textrm{np},
\end{align}
where
\begin{align}
	\label{Mstability}
	\renewcommand{\arraystretch}{1.7}
	\setlength{\arraycolsep}{1pt} % adjusting column spacing
	\resizebox{1.0\linewidth}{!}{
		$M_\textrm{np}=\omega_0
		\left(\begin{array}{@{\hspace{0pt}}ccc@{\hspace{0pt}}}
			-2\bar\kappa & -i2 g_rg_{cr} & i2 g_rg_{cr} \\
			i4 g_rg_{cr} & i2(g_r^2+g_{cr}^2-1)-2\bar\kappa & 0 \\
			-i4g_rg_{cr} & 0 & i2(1-g_r^2-g_{cr}^2)-2\bar\kappa
		\end{array}\right)$
	}
\end{align}
and	$Y_\textrm{np}=\omega_0\left(0, i2g_r g_{cr},- i2g_r g_{cr}\right)^\intercal$. The solution for the steady state is given by
\begin{align}
	\renewcommand{\arraystretch}{1.2}
	\textbf{s}_s=-M_\textrm{np}^{-1}Y_\textrm{np}.
\end{align}
Explicitly, the excitation number of the oscillator for the steady state reads
\begin{align}
	\label{eq: ada}
	\braket{a^\dagger a}_s=\frac{2 g^4 \varepsilon ^2}{g^4 \left(1-\varepsilon ^2\right)^2-2 g^2 \left(1+\varepsilon
		^2\right)+\bar\kappa ^2+1}.
\end{align}
The number of excitations diverges at the boundary of the NP, and the divergence exhibits a power law, whose critical exponents are examined in the next section.

%-----------------------------------------------
%-----------------------------------------------

\subsection{Superradiant phase}\label{sec:Superradiant Phase}
Let us now examine the quantum dynamics and fluctuations in the SP. Following the method in Ref.~\cite{Hwang2018PRA}, we derive an effective master equation in the SP by applying a displacement unitary transformation $\mathcal{D}[\alpha]=\exp[\alpha a^\dagger -\alpha^* a]$ to the master equation~(\ref{eq:d_rho}), followed by the SW transformation. See details in Appendix ~\ref{App:Hsp_eff}. The effective master equation for the SP becomes
\begin{equation}
	\dot \rho_\text{sp}=-i[H_q,\rho_\text{sp}]+\kappa D[a]\rho_\text{sp},
\end{equation}
where the quadratic effective Hamiltonian $H_q$ reads
\begin{equation}
	H_q=P a^\dagger a +Q  a a  +Q^* a^\dagger a^\dagger.
\end{equation}
The expressions for the coefficients $P$ and $Q$ are quite involved and they are given in Eqs.~(\ref{eq:P}) and (\ref{eq:Q}) in Appendix ~\ref{App:Hsp_eff}.

The dynamics of the mean value of the oscillator operator is determined by $\dot{\textbf{a}}=L_\textrm{sp} \textbf{a}$ with
\begin{align}
	\label{SPstability}
	\renewcommand{\arraystretch}{1.2}
	L_\textrm{sp}=\left(\begin{array}{cc}-i P-\kappa& -i2Q^* \\i2Q& iP-\kappa\end{array}\right).
\end{align}
The eigenvalues $l_\textrm{sp}$ of $L_\textrm{sp}/\omega_0$ are given by
\begin{align}
	\label{eq:l_sp}
	l^{\pm}_\textrm{sp}=-\bar\kappa \pm \frac{1}{\omega_0}\sqrt{4 |Q|^2 - P^2}.
\end{align}
The stability condition of the SP, $\mathrm{Re}[l_\textrm{sp}^\pm]<0$, agrees well with the stable area for the SP as shown in Fig.~\ref{fig:phasediagram}(b). The real part (blue dashed curve) and imaginary part (orange dash-dotted curve) of  $ l_\textrm{sp}^+ $ are shown in Fig.~\ref{fig:Lnpsp}(a), where $g$ goes along the blue dashed line in Fig.~\ref{fig:phasediagram}(c). Similar to the NP, the real part of $l_\textrm{sp}^+$ approaches zero at phase boundaries of the SP, leading to a vanishing ADR. In Fig.~\ref{fig:Lnpsp}(a), we observe a range of $g$, $1.51 \lesssim g \lesssim 2.55$, where the real part of both $l_\textrm{sp}^+$ and $l_\textrm{np}^+$ are negative, which indicates the emergence of the bistable phase.

\begin{figure}[tpb!]
	\hspace{0.25cm}\includegraphics[width=0.95\linewidth]{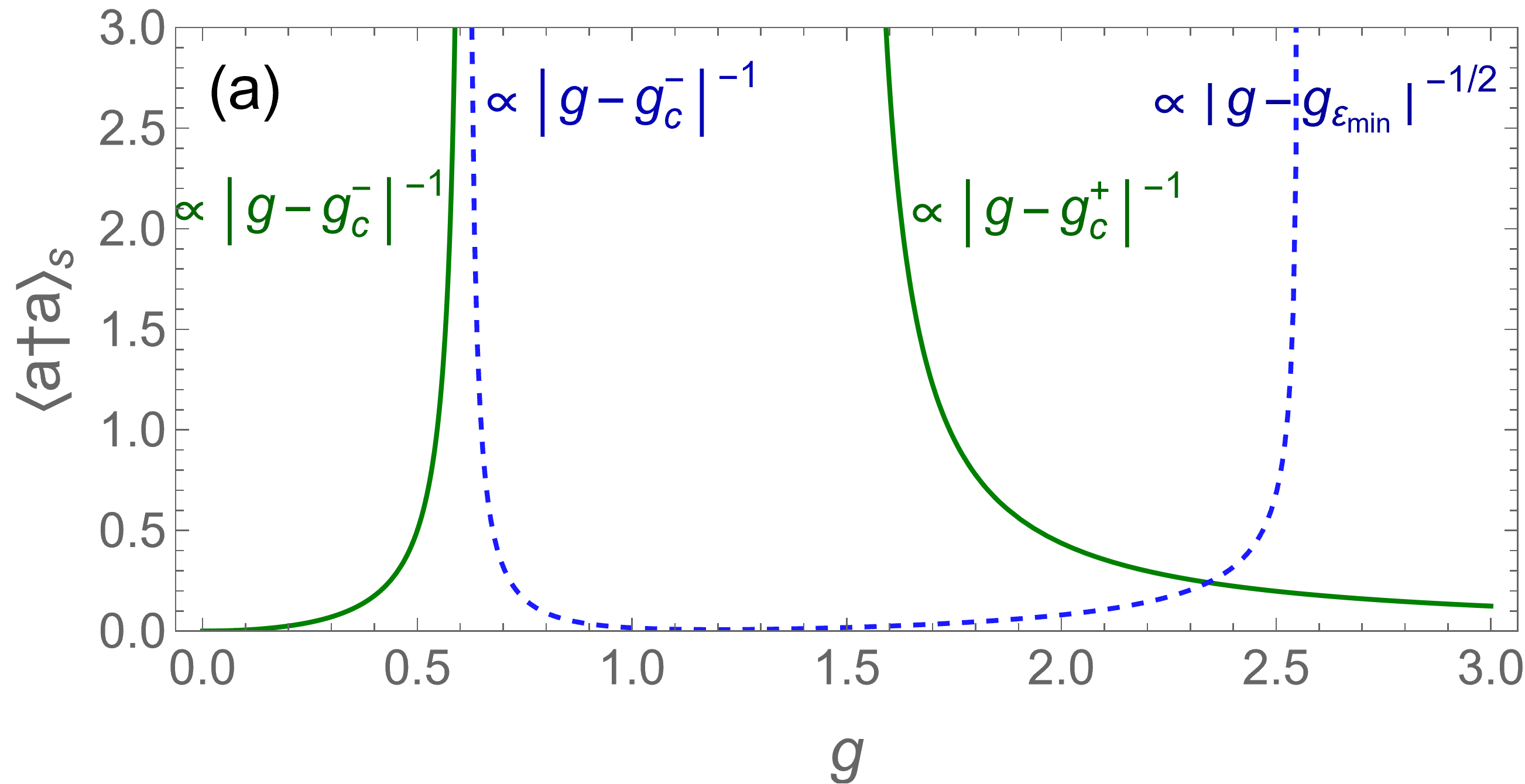}
	\vspace{0.3cm}

	\includegraphics[width=0.975\linewidth]{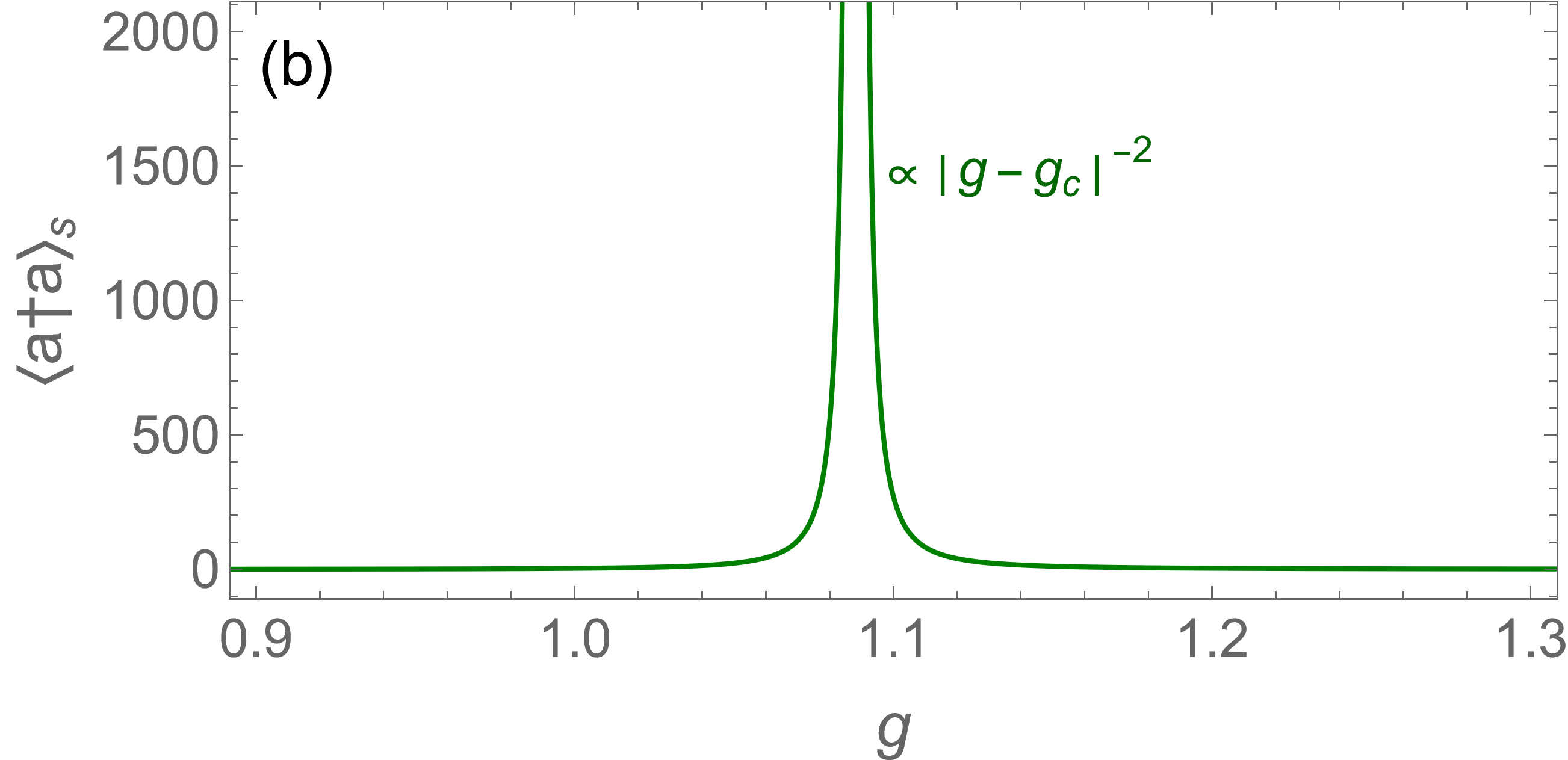}
	\caption{Critical scaling of the oscillator population. The oscillator population for the NP (green solid) and the SP (blue dashed) as a function of $g$ is plotted as one moves (a) along the blue dashed line and (b) along the red dotted line in Fig.~\ref{fig:phasediagram}(c). The latter represents the critical behavior across the tricritical point. The scaling exponent near the critical coupling strength is presented in each panel. Here, $\bar\kappa=0.5$.
	}
	\label{fig:photnum}
\end{figure}

By solving the equation of motion of the second moments, we find that the excitation number of the oscillator in the SP is given by
\begin{align}
	\label{eq:ada_sp}
	\braket{a^\dagger a}_s= \frac{2|Q|^2}{P^2-4 |Q|^2+ \kappa^2}.
\end{align}
As shown in Fig.~\ref{fig:photnum}(a), the excitation number diverges in the SP at the phase boundaries.

%--------------------------------------
%--------------------------------------

\section{Universality of the tricritical point} \label{sec: Universality}
Using the quantum solutions derived in the previous section, we now examine the universality for the DPTs in the anisotropic open QRM. The closing of the Liouvillian gap, thus the vanishing ADR, and the diverging oscillator excitation number are the characteristic features of a second-order DPT driven by a Markovian bath~\cite{TorrePRA2013,Hwang2018PRA}. These features can be clearly seen in Figs.~\ref{fig:Lnpsp}(a) and  \ref{fig:photnum}(a).

First, let us focus on the critical curve $g_c^-(\varepsilon_{\text{min}}<\varepsilon<\varepsilon_{\text{max}},\bar\kappa)$ at which a second-order phase transition occurs. On both sides of the critical curve, the ADR vanishes as
\begin{align}
	\label{eq:ADR}
	\kappa_\textrm{ADR}=\mathrm{Re}[l_\textrm{np (sp)}^+]\propto|g-g_c^{-}(\varepsilon,\bar\kappa)|^{\nu_\textrm{ADR}},
\end{align}
with $\nu_\textrm{ADR}=1$, and the oscillator excitation number diverges as
\begin{align}
	\label{eq:nu_x}
	\braket{a^\dagger a}_s\propto|g-g_c^{-}(\varepsilon,\bar\kappa)|^{-\nu_x},
\end{align}
with the exponent $\nu_x=1$. Therefore, the second-order DPT along the critical curve $g_c^-$ belongs to the same universality class as the symmetric open QRM~\cite{Hwang2018PRA} and open Dicke model~\cite{TorrePRA2013}.

Second, we have two additional critical curves $g_c^+(\varepsilon,\bar\kappa)$ where a first-order phase transition occurs due to the bistability. In equilibrium, correlation functions and fluctuations typically do not diverge at the first-order phase transition~\cite{Nishimori2010}. In stark contrast, in our nonequilibrium model, the number of oscillator excitations (a fluctuation of the order parameter) diverges with a power-law at the first-order phase transition. The exponents at the first-order phase boundary of the NP, $g_c^+(\varepsilon,\kappa)$, are the same as those of the second-order phase transition, namely, $\nu_\textrm{ADR}=\nu_x=1$. At the first-order boundary of the SP, on the other hand, we find a new set of exponents $\nu_\textrm{ADR}=\nu_x=1/2$ [see Figs.~\ref{fig:Lnpsp}(a) and  \ref{fig:photnum}(a)]. Our finding shows that the nonequilibrium first-order DPTs driven by a Markovian bath could show drastically different critical properties from its equilibrium counterpart \cite{Nishimori2010}. We note that the denominator of the excitation number in Eq.~(\ref{eq: ada}) has the same roots [i.e. $g_c^\pm(\varepsilon,\bar\kappa)$] as the eigenvalue of $l_{\text{np}}^+$ in Eq.~(\ref{eq:lnp}). The same holds for the SP case, as seen in Eqs.~(\ref{eq:ada_sp}) and (\ref{eq:l_sp}). Therefore, the ADR vanishes at the critical coupling strengths $g_c^\pm$ and $g_{\varepsilon_\text{min}}$ with the same critical exponent that governs the divergence of excitation number [see Figs.~\ref{fig:Lnpsp}(a) and  \ref{fig:photnum}(a)].

The critical exponents of the tricritical point are typically different from those of second-order phase transition~\cite{Nishimori2010}; therefore, it is interesting to examine the criticality of the tricritical point $g_c^{-}(\varepsilon_{\text{min (max)}},\bar\kappa)$. We find that the critical exponents depend on the angle at which one crosses the critical point. The novel critical behaviors along the line $g_{cr}= g_r \varepsilon_\text{min}$ are presented in Figs.~\ref{fig:Lnpsp}(b) and \ref{fig:photnum}(b), which show that the critical exponents are given as $\nu_\textrm{ADR}=\nu_x=2$. We note that $g_{cr} = g_r \varepsilon_{\min}$ is actually the tangent line of the NP boundary at the tricritical point.
Along this line, both sides of the tricritical point are the NP and there is no phase transition.
In fact, if one moves along any line that is tangent to the phase boundary of the NP, we find the identical critical behaviors shown in Figs.~\ref{fig:Lnpsp}(b) and  \ref{fig:photnum}(b).
To understand this set of higher critical exponents, one can see the real part of $\l_\text{np}^+$ shown in Fig.~\ref{fig:Lnpsp}(b). Near the critical value $g_c \approx 1.1$, the top of the parabolic-like curve ($\mathrm{Re}[l_{\text{np}}^+]$) touches $0$, therefore presenting the critical exponent $\nu_\textrm{ADR} = 2$.
But if $\mathrm{Re}[l_{\text{np}}^+]$ becomes greater than zero, indicating that $g$ goes across the critical point $g_c$ and the system goes though a phase transition, the critical exponent will turn to $\nu_{\text{ADR}}=1$ as in the case shown in Fig.~\ref{fig:Lnpsp}(a).
This showcases the peculiarity of the tangent line along the convex boundary of the NP, which leads to a set of higher scaling exponents $\nu_{\text{ADR}}=\nu_x=2$.
Furthermore, we verify that this property is also applicable for the generalized open Dicke model \cite{ETH2018PRL}. Our analysis demonstrates that the non-monotonic phase boundary due to the competition between the dissipation and the coherent interaction gives rise to various ways of crossing the critical points with different critical exponents. It would be interesting to investigate whether these novel critical scalings could be used to improve the critical sensing schemes based on the open QRM~\cite{Ilias2022PRXQuantu, Ilias2023arXiv}.

\section{Implementation} \label{sec: Implementation}
Reference~\cite{Hwang2018PRA} proposed to use a trapped-ion setup involving two ions to implement the open QRM. Here, we briefly review the proposal and discuss how the same setup could be used to realize the anisotropic open QRM.

Consider a mixed species ion pair, $^{9}{\rm Be^+-^{24}}{\rm Mg}^+$~\cite{Lin2013PRL,Tan2015Nature}, in a linear Paul trap. The common center-of-mass mode can be used as the oscillator realizing the open QRM, while all other vibration modes are significantly detuned. The hyperfine states of the $^{9}{\rm Be^+}$ ion, specifically $\ket{F=2,m_F=0}$ and $\ket{F=1,m_F=1}$, form a qubit that can be coupled to the motional mode through coherent stimulated Raman transitions~\cite{Monroe1995PRL}. By applying two lasers to the $^{9}{\rm Be^+}$ ion, with detunings $\delta_1$ and $\delta_2$, we drive the red- and blue-sideband transitions. The intensity of the two lasers can be controlled independently to realize different Rabi frequencies $\Omega_1^d$ and $\Omega_2^d$. In the interaction picture with respect to the bare qubit and oscillator dynamics, followed by a rotating wave approximation, the interaction Hamiltonian between the oscillator and the qubit in the Lamb-Dicke limit can be expressed as $H_I=-\eta_{\rm LD}\Omega_1^d /2 ( \sigma_+ a^\dagger \mathrm{e}^{i\delta_1 t}+\mathrm{H.c.})-\eta_{\rm LD}\Omega_2^d  /2( \sigma_+ a \mathrm{e}^{i\delta_2 t}+\text{H.c.})$, where $\eta_{\rm LD}$ represents the Lamb-Dicke parameter. Moving to the rotating frame of $H_\text{rot}=\sigma_{z} (\delta_1+\delta_2)/4 + a^{\dagger} a  (\delta_1-\delta_2)/2 $, the interaction Hamiltonian $H_I$ turns into the form of our Rabi Hamiltonian (\ref{eq:Hamiltonian}). In this frame, the parameters of the Rabi Hamiltonian are given by $\omega_0=(\delta_1-\delta_2)/2$, $\Omega=(\delta_1+\delta_2)/2$, $\lambda_{cr}=\eta_{\rm LD}\Omega_1^d /2$, and $\lambda_{r}=\eta_{\rm LD}\Omega_2^d /2$~\cite{Pedernales2015SciRep,Puebla2017PRL}. Therefore, simply by controlling the relative intensities of the Raman lasers driving blue- and red-sideband transitions, one could realize the anisotropic Rabi model. The dissipation to the oscillator can be achieved by using the sympathetic cooling~\cite{Lin2013PRL,Tan2015Nature} of the center-of-mass mode with the second ion, $^{24}{\rm Mg}^+$. Previous experimental studies have successfully demonstrated the  sympathetic cooling for the in-phase mode using $^{24}{\rm Mg}^+$ ions~\cite{Lin2013PRL,Tan2015Nature}. Moreover, it has been experimentally demonstrated that coherent bosonic states with a large number of excitations can be generated for the motional modes in ion-traps using sideband transitions and dissipation~\cite{Kienzler:2015gc,Kienzler2017PRL,Behrle2023PRL}.

\section{Conclusion}  \label{sec: Conclusion}

In conclusion, we have investigated a rich steady-state phase diagram of the anisotropic open QRM. It features a number of interesting critical phenomena resulting from the competition between the anisotropic interaction and the dissipation. First, there are two critical ratios, $\varepsilon_\textrm{min}$ and $\varepsilon_\textrm{max}$, between the interaction strengths of the counterrotating term and the rotating term, beyond and below which no DPT occurs. In such cases, the Markovian bath brings the system to a trivial steady state, regardless of the strength of the coherent interaction.
Second, in between these critical ratios, a second-order DPT occurs where the NP becomes unstable and the SP with a broken symmetry emerges. Strikingly, upon increasing further the coherent interaction strength beyond the second-order DPT, the NP becomes stable again, leading a bistable phase of the NP and the SP. The boundaries of the bistable phase are characterized by a first-order DPT. Moreover, a tricritical point appears where the critical lines for the second-order and the first-order DPTs meet. Third, the tricritical point features a different set of critical exponents. It is an interesting question to see whether the novel scaling behavior of the tricritical point could be harnessed for the critical quantum sensing protocol based on the open QRM~ \cite{Ilias2022PRXQuantu, Ilias2023arXiv}.
Our work demonstrates that interesting nonequilibrium critical phenomena such as multicriticality and bistability can be explored in a finite-component systems with a controlled coherent interaction and damping, realized by two trapped ions.

\textit{Note added.} We became aware of a related work \cite{Li2023arXiv} on the anisotropic open QRM, which appeared during the review process, where the stability of the spin up state and its role on the phase diagram in the finite frequency regime is investigated and the adiabatic elimination of spin variables is also presented.

\begin{acknowledgments}
G.L. and M.-J.H. acknowledge support from Kunshan Municipal Government research funding, and the Innovation Program for Quantum Science and Technology Grant No. 2021ZD0301602. K.K. acknowledges funding from the European Union's Horizon 2020 research and innovation programme under the Marie Sklodowska-Curie Grant Agreement No. 713729. This work was supported by the EU project C-QuENS (Proposal No. 10113539).
\end{acknowledgments}

%\newpage
%\clearpage

\appendix
\setcounter{figure}{0}
\renewcommand{\thefigure}{A\arabic{figure}}

\section{Stability Analysis}\label{App:Stability Analysis}
Here, we provide a stability analysis for the mean-field solutions. To this end, we consider small deviations from the mean-field solutions given in Eqs.~(\ref{eom1}-\ref{eom3}): $\braket{a} \mapsto \alpha + \delta \alpha$, $\braket{\sigma_+} \mapsto s_+ + \delta s_+$, and $\braket{\sigma_z} \mapsto s_z + \delta s_z$. Considering only the first-order terms for the fluctuation, we obtain that
\begin{align}
	&\dfrac{d}{dt}\delta \alpha  = -i (\omega_0 - i \kappa) \delta \alpha  + i (\lambda_r \delta s_- + \lambda_{cr} \delta s_+),\\[0.3ex]
	&\resizebox{1\linewidth}{!}{$\dfrac{d}{dt}\delta s_+ = i \Omega \delta s_+ + i (\lambda_r \delta \alpha ^* + \lambda_{cr} \delta \alpha ) s_z + i(\lambda_r \alpha^* + \lambda_{cr} \alpha) \delta s_z,$}\\[0.3ex]
	&\resizebox{1\linewidth}{!}{$\dfrac{d}{dt} \delta s_z = -4 \mathrm{Im}\left[(\lambda_r \alpha + \lambda_{cr} \alpha^*)\delta s_+ + (\lambda_r \delta \alpha + \lambda_{cr} \delta \alpha ^*) s_+\right].$}
\end{align}
By introducing $\delta \alpha  \equiv  \delta x + i \delta y$ and $\delta s_+ \equiv (\delta s_x + i \delta s_y)/2$, and using the condition for the spin conservation $s_x^2 +s_y^2 +s_z^2 = 1$, the dynamical equation becomes
\begin{widetext}
	\begin{equation}
		\renewcommand{\arraystretch}{1.2}
		\dfrac{d}{dt}
		\begin{pmatrix}
			\delta x \\
			\delta y \\
			\delta s_x \\
			\delta s_y \\
		\end{pmatrix} =
		\begin{pmatrix}
			- \kappa & \omega_0 & 0 & \lambda(1-\varepsilon)/2 \\
			-\omega_0 & - \kappa & \lambda(1 + \varepsilon)/2 & 0 \\
			0 &  2\lambda (1-\varepsilon)s_z & -2\lambda(1-\varepsilon)y \frac{s_x}{s_z} & -\left(  2\lambda(1-\varepsilon)y \frac{s_y}{s_z} + \Omega \right) \\
			2\lambda(1+\varepsilon) s_z & 0 & - \left(2\lambda(1+\varepsilon) x \frac{s_x}{s_z} - \Omega\right) & -2\lambda(1+\varepsilon) x \frac{s_y}{s_z}
		\end{pmatrix}
		\begin{pmatrix}
			\delta x \\
			\delta y \\
			\delta s_x \\
			\delta s_y \\
		\end{pmatrix}.
	\end{equation}
\end{widetext}

The eigenvalues $\mu$ of the above matrix determine the stability of the mean-value solutions. Considering the limit of $ \Omega/\omega_0 \rightarrow \infty $,  we find the characteristic equation up to the leading-order term of $\Omega/\omega_0$, which reads $A{\omega_0}^2 + 2 \bar{\kappa}B\omega_0 \mu + B \mu^2=0$, where
\begin{align}
	\label{eq:A2}
	A &= 1+ \bar{\kappa}^2 + 4 g^2 (1+\bar{\kappa}^2)[(-1+\varepsilon)^2 \bar{y}^2 + (1+\varepsilon)^2\bar{x}^2]\nonumber \\
	&\quad + 2g^2 (1+\varepsilon^2) s_z +g^4  (-1+\varepsilon^2)^2 [s_z^2 + 4s_z(\bar{x}^2+\bar{y}^2)],\\
	B &= 1 + 4 g^2\left[(-1+\varepsilon)^2\bar{y}^2 + (1+\varepsilon)^2\bar{x}^2 \right].
\end{align}
Note that we have used the expressions of $s_x$ and $s_y$ given in Eq.~(\ref{eq:s_x_y}) and the renormalized parameters to rewrite the coefficients given above. Finally, we obtain the eigenvalue
\begin{equation}
	\mu / \omega_0 = - \bar{\kappa} \pm \sqrt{\bar{\kappa}^2 - A/B}.
\end{equation}
The stability condition $\text{Re}(\mu) < 0$ reduces to $A>0$, because $B$ is always positive.

\section{Stability of the Nontrivial Solution with $s_z^+$ }\label{App:SP with s_z^+}
\begin{figure}[hpt!]
	\includegraphics[width=0.60\linewidth]{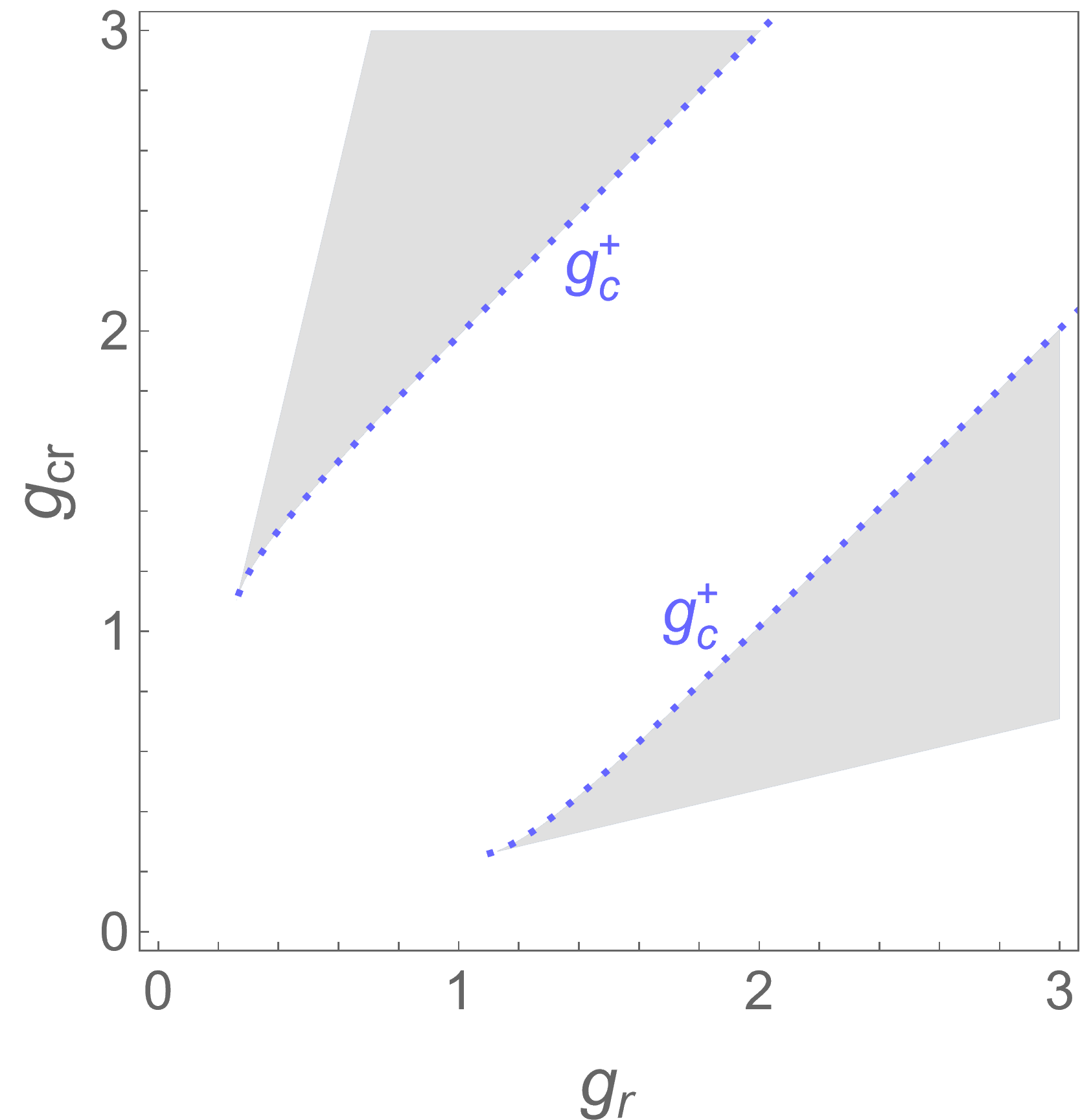}
	\caption{Physical region for the nontrivial solution with $ s_z^+ $.
		The blue dotted curves represent the critical coupling strength $g_c^+(\varepsilon,\bar{\kappa})$. The gray area shows the physical allowed region with $-1<s_z^+<1$. Here we set $\bar\kappa=0.5$.
	}
	\label{fig:phase_SP_s_z^+}
\end{figure}

Here, we check the stability of the nontrivial solution $s_{z}=s_{z}^+\equiv -[(1+\varepsilon^2)+\sqrt{4\varepsilon^2-\bar{\kappa}^2(1-\varepsilon^2)^2}]/(1-\varepsilon^2)^2g^2$ for Eq.~(\ref{eq:det_Lcl_0}). By substituting $ s_{z}^+ $, $x$ [Eq.~(\ref{result_x})], and $y$ [Eq.~(\ref{result_y})] into the coefficient $A$ given in Eq.~(\ref{eq:A2}), we find that the stability condition $A>0$ is equivalent to
\begin{equation}
g<\sqrt{\frac{(1+\varepsilon^2)+\sqrt{4\varepsilon^2-\bar{\kappa}^2(1-\varepsilon^2)^2}}{(1-\varepsilon^2)^2}}.
\label{eq:B1}
\end{equation}
We notice that the right hand side of the inequality is the same as the value of $g_c^+(\varepsilon,\bar{\kappa})$ given in Eq.~(\ref{eq:g_c}). The critical coupling strength $g_c^+(\varepsilon,\bar{\kappa})$ is shown by the blue dotted curves in Fig.~\ref{fig:phase_SP_s_z^+}. The solution for the $s_z$ is meaningful only when $-1<s_z^+<1$. We plot the region $-1<s_z^+<1$ in gray in Fig.~\ref{fig:phase_SP_s_z^+}. In this region, the condition~(\ref{eq:B1}) is not satisfied, therefore, the physical solution of the $s_z^+$ is unstable.

%---------------------------------
%---------------------------------

\section{Adiabatic elimination and numerical simulations for the semiclassical equations of motion} \label{App:simulation}

We rewrite the mean-field equations of motion Eqs.~(\ref{eom1}-\ref{eom3}) using a rescaled time $\bar{t}\equiv \omega_0 t$ as
\begin{align}
	\label{EOM1} \dfrac{d \bar{\alpha} }{d\bar{t}}&=-i (1-i\bar\kappa) \bar{\alpha}+ i (\gr s_-+ \gcr s_+),\\
	\label{EOM2}\dfrac{1}{\eta} \dfrac{d s_+}{d\bar{t}}&= i s_+ + i (\gr \bar{\alpha}^*+\gcr \bar{\alpha})s_z,\\
	\label{EOM3}\dfrac{1}{\eta} \dfrac{d s_z}{d\bar{t}}&=i (\gr \bar{\alpha}+\gcr \bar{\alpha}^*)s_+ -i(\gr \bar{\alpha}^*+ \gcr \bar{\alpha})s_-,
\end{align}
where $\eta$ is the frequency ratio $\Omega/\omega_0$. In the thermodynamic limit $\eta \rightarrow \infty$, the left-hand-side of Eqs.~(\ref{EOM2}) and (\ref{EOM3}) vanish, giving rise to
\begin{align}
	\label{s_plus} s_+ &= -(\gcr \bar{\alpha}+ \gr \bar{\alpha}^*) s_z,\\
	\label{s_z} s_z&= \pm \dfrac{1}{\sqrt{1+4|\gcr \bar{\alpha}+ \gr \bar{\alpha}^*|^2}}.
\end{align}
Namely, the spin dynamics instantaneously follows the changes of the cavity dynamics in this limit. By putting Eqs.~(\ref{s_plus}) and (\ref{s_z}) back to Eq.~(\ref{EOM1}), we obtain a nonlinear equation of motion for the cavity field,
\begin{align}
	\label{EOM4}
	 \dfrac{d \bar{\alpha} }{d\bar{t}}&=-i (1-i\bar\kappa) \bar{\alpha} \mp i \dfrac{(\gr^2+\gcr^2) \bar{\alpha}+ 2\gr\gcr \bar{\alpha}^*}{\sqrt{1+4|\gcr \bar{\alpha}+ \gr \bar{\alpha}^*|^2}}.
\end{align}
The stability analysis of Eq.~(\ref{EOM4}) exactly reproduces the phase diagram depicted in Fig.~\ref{fig:phasediagram}(c).

\begin{figure}[tpb!]
	\includegraphics[width=0.49\linewidth]{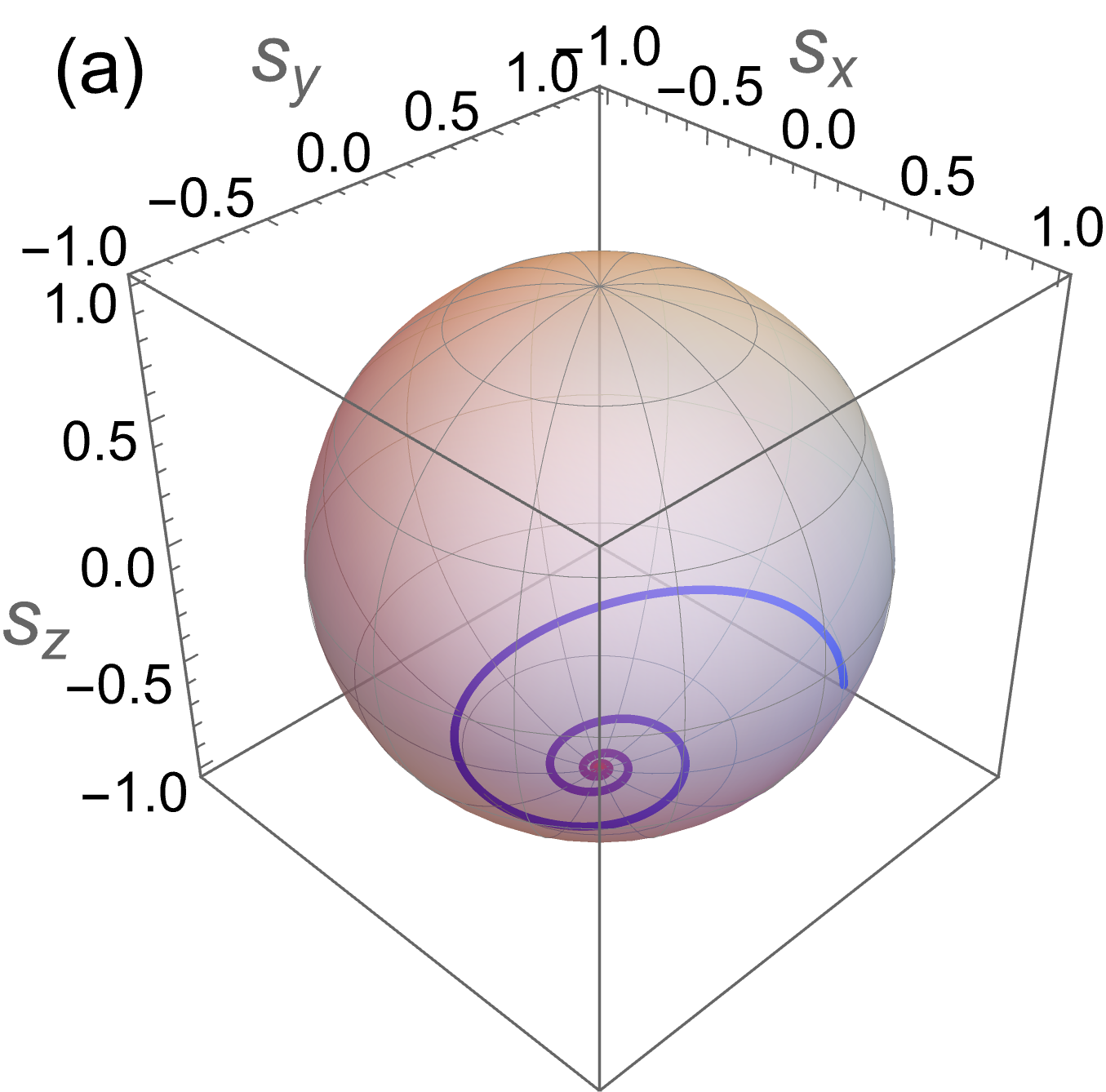}
	\includegraphics[width=0.49\linewidth]{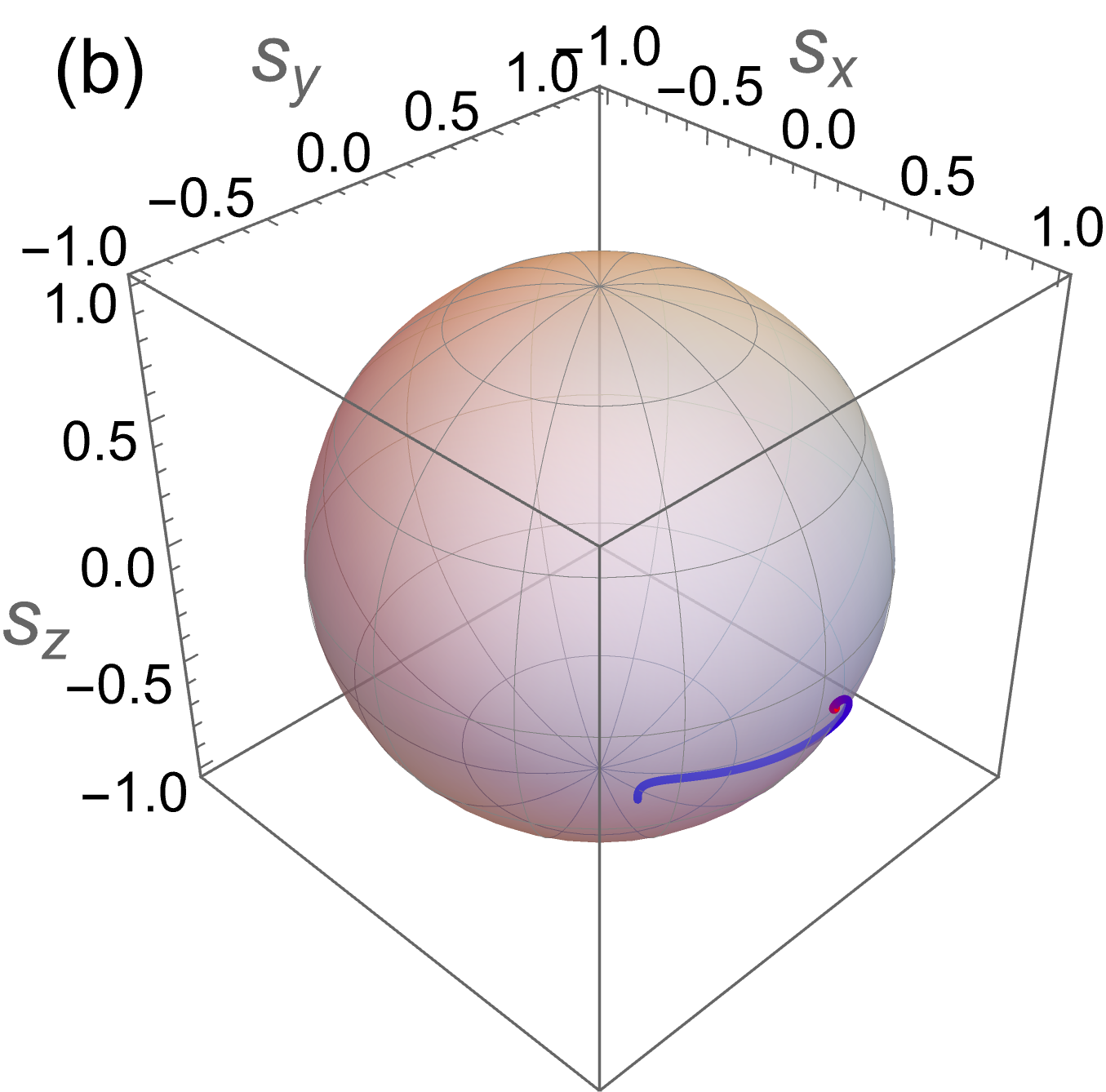}
	\vspace{0.1cm}
	\includegraphics[width=0.49\linewidth]{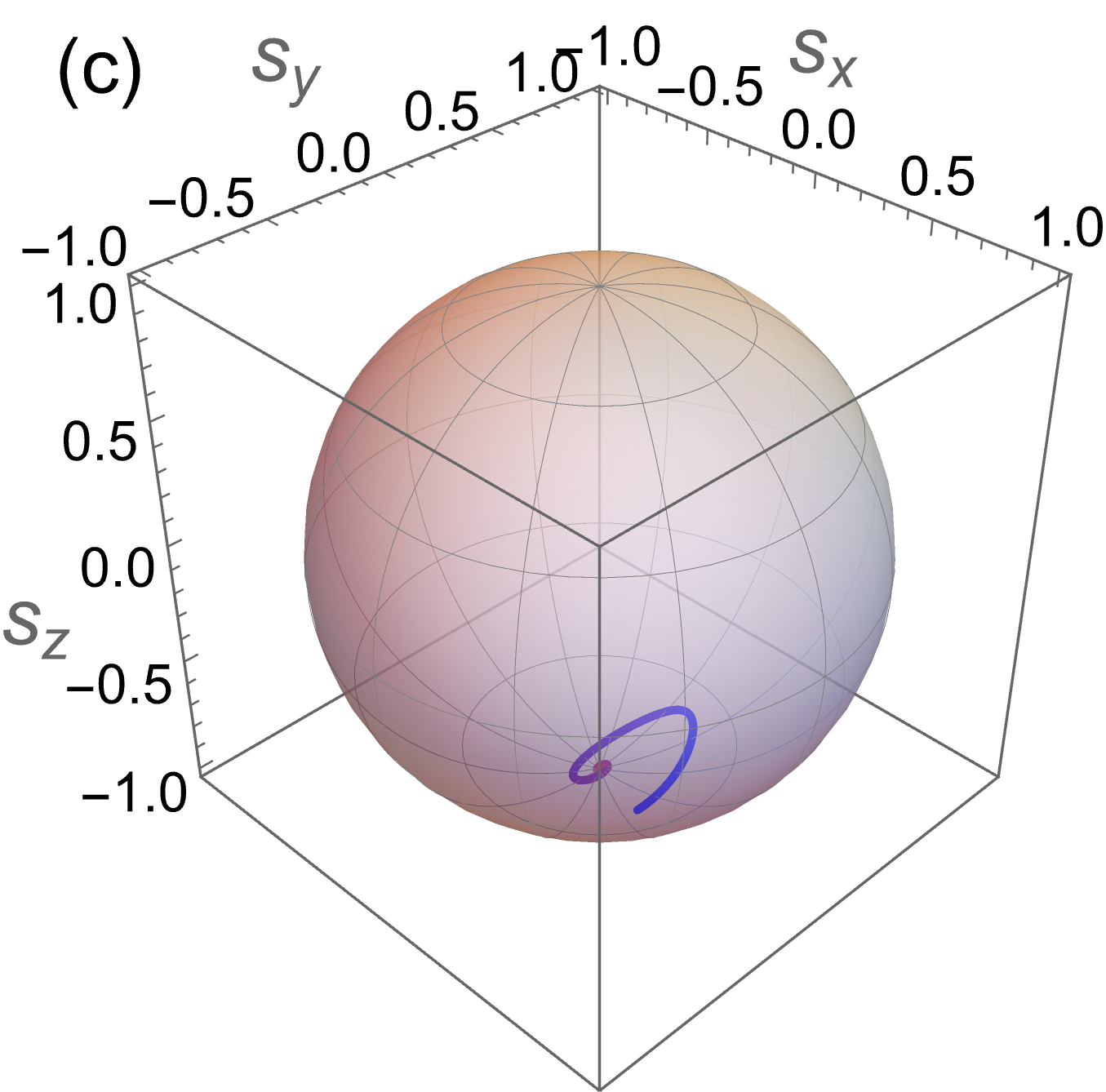}
	\includegraphics[width=0.49\linewidth]{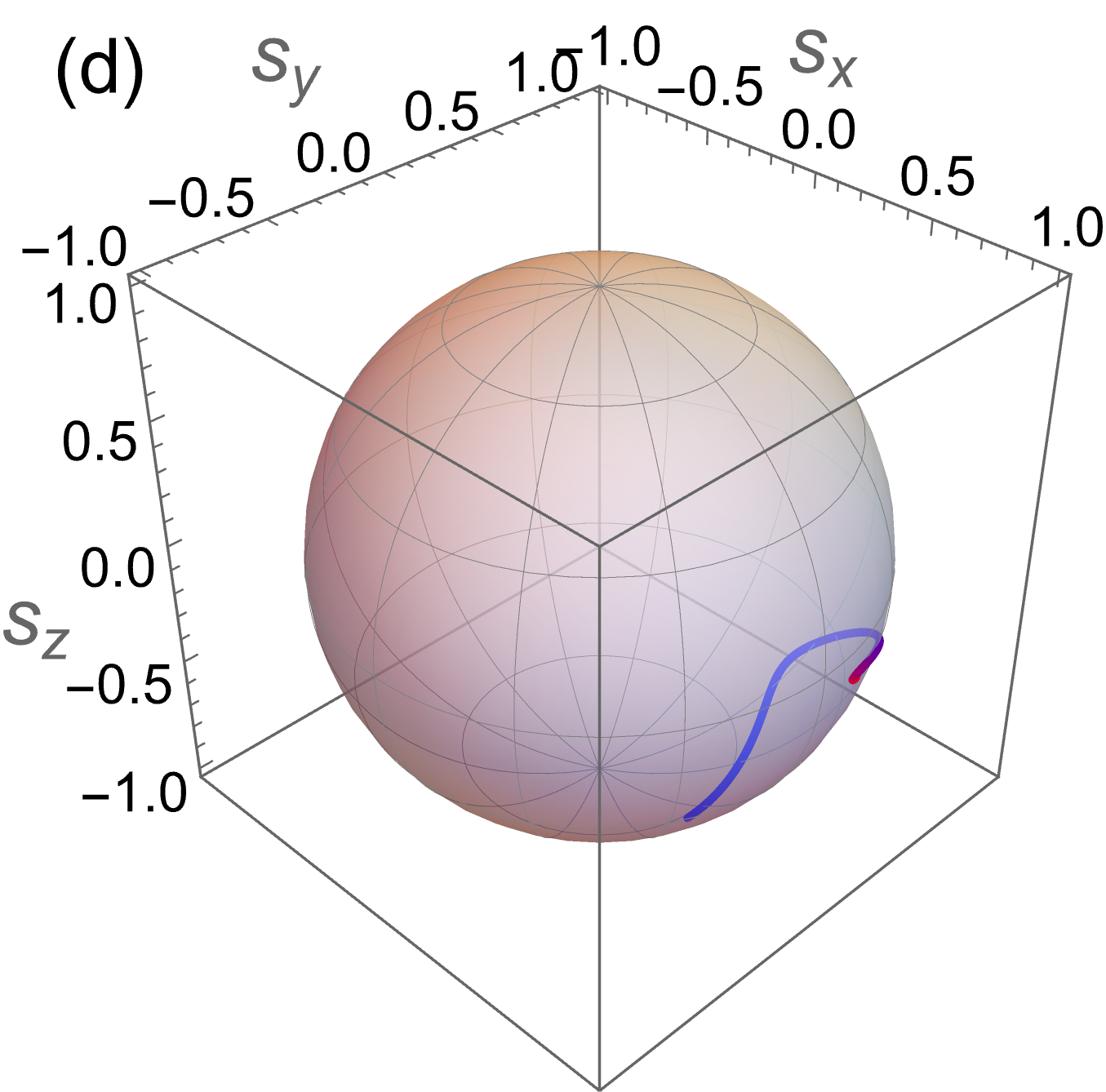}
	\caption{Trajectories of the spin on the Bloch sphere. Here, we set $\bar\kappa=0.5$. (a) Normal phase. For $g_{r}=0.3$, $g_{cr}=2$, and the initial state with $\bar{\alpha}=0.3+0.3i$ that corresponds to $\{s_x,s_y,s_z\}\approx \{0.69, 0.51, -0.50\}$, the spin evolves to the normal steady state. (b) Superradiant phase. For the case with $g_{r}=1$, $g_{cr}=1.5$  and the initial state with $\bar{\alpha}=0.05-0.05i$ that corresponds to $\{s_x,s_y,s_z\}\approx \{0.24, -0.05, -0.97\}$, the spin evolves to the superradiant steady state. (c, d) Bistable phase. We set $g_{r}=1$, $g_{cr}=2.1$. The initial state with $\bar{\alpha}=0.05-0.05i$ that corresponds to $\{s_x,s_y,s_z\}\approx \{0.29, -0.10, -0.95\}$ evolves to the normal steady state, while the initial state with $\bar{\alpha}=0.1-0.05i$ that corresponds to $\{s_x,s_y,s_z\}\approx \{0.52, -0.09, -0.85\}$ evolves to the superradiant steady state.
	}
	\label{fig:trajectories}
\end{figure}

Furthermore, we numerically solve the mean-field equation of motion, Eq.~(\ref{EOM4}), for various initial conditions to corroborate the absence of the nonstationary steady-state solutions that are reported for the open Dicke model in Ref.~\cite{Stitely2020PRR}. In Fig.~\ref{fig:trajectories}, we illustrate the trajectory of the spin on the Bloch sphere for various initial state with $s_z<0$. In the NP, the spin evolves into the normal steady state with $s_z=-1$ [see Fig.~\ref{fig:trajectories}(a)]. In the SP, on the other hand, the system evolves into the superradiant steady state with $-1<s_z<0$ [see Fig.~\ref{fig:trajectories}(b)]. In the bistable phase, the system evolves either to the normal or superradiant steady state, depending on the initial states [see Fig.~\ref{fig:trajectories}(c) and \ref{fig:trajectories}(d), respectively]. Our numerical results suggest that there are basins of attraction in the Bloch sphere for both NP and SP solutions in the bistability phase, whose detailed characterization is left for future study. Note that for the initial states with $s_z>0$, the system always evolves into the state with $s_z=+1$ (not shown). The initial condition $s_z=0$ corresponds to $\bar{\alpha}\rightarrow\infty$, which is not a physically meaningful initial state.
%---------------------------------
%---------------------------------

\section{Numerical simulations for finite frequency ratio} \label{App:quantum simulation}
\begin{figure}[tpb!]
	\includegraphics[width=0.49\linewidth]{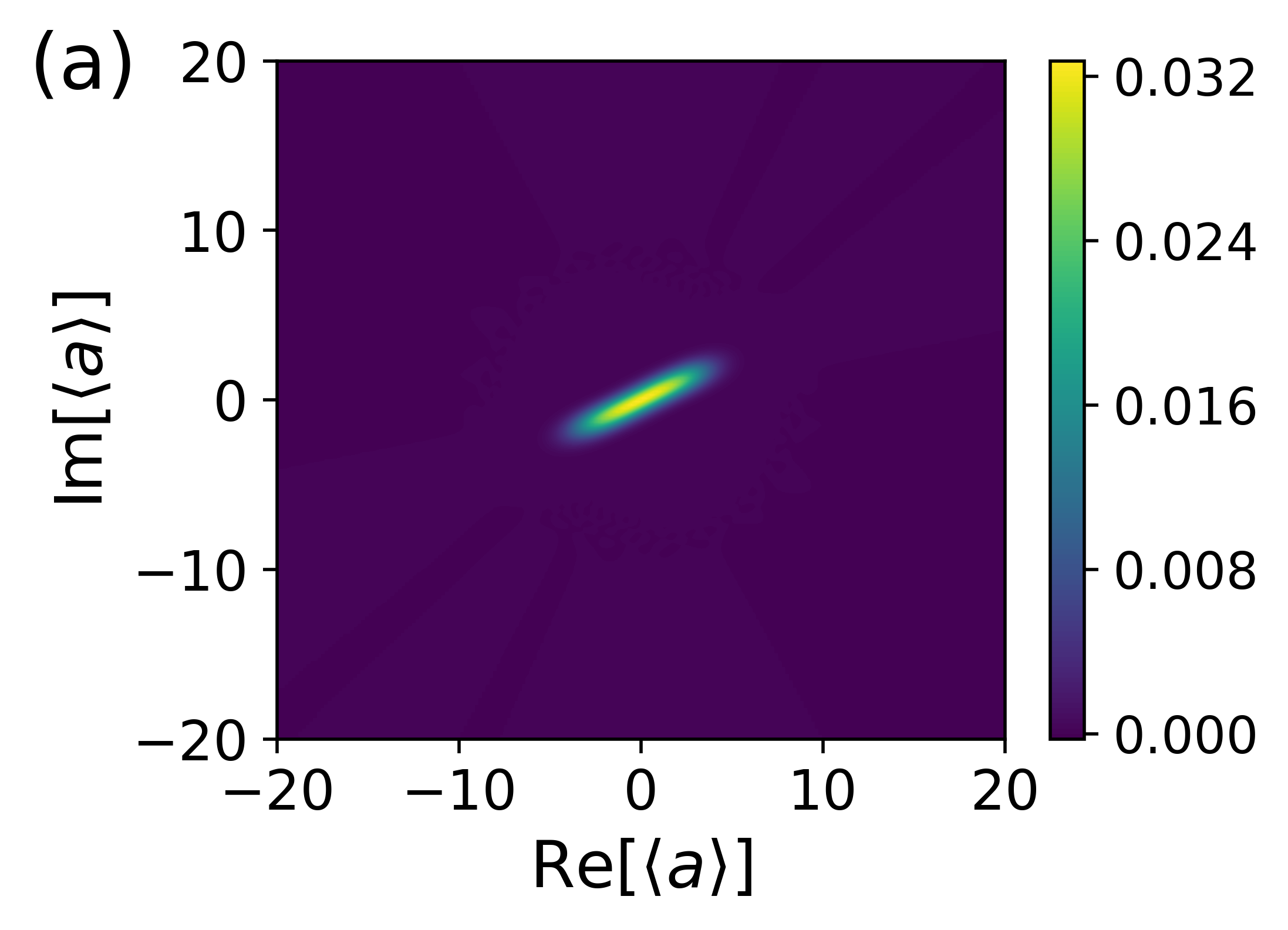}
	\includegraphics[width=0.49\linewidth]{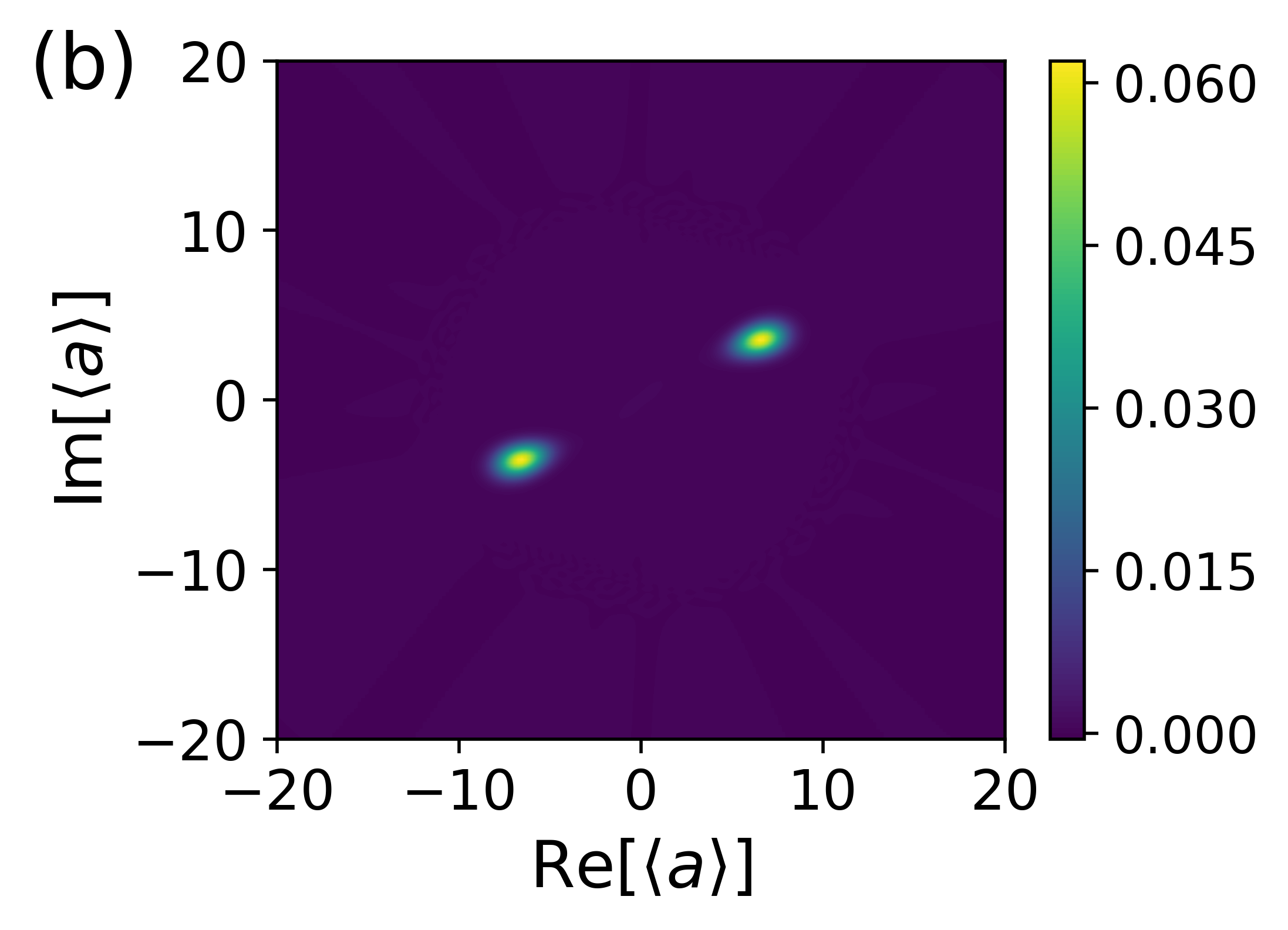}
	\vspace{0.1cm}
	\includegraphics[width=0.49\linewidth]{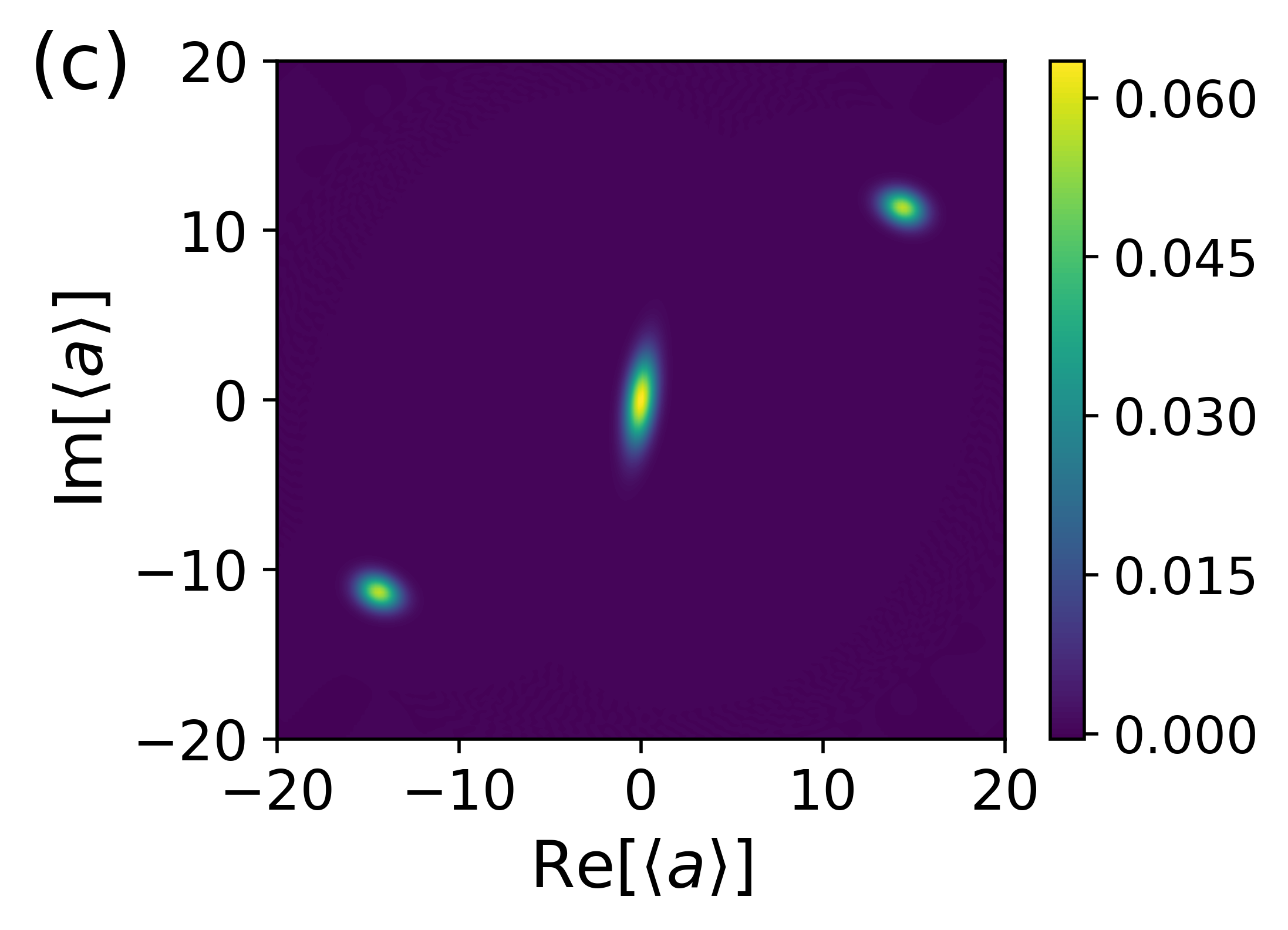}
	\includegraphics[width=0.49\linewidth]{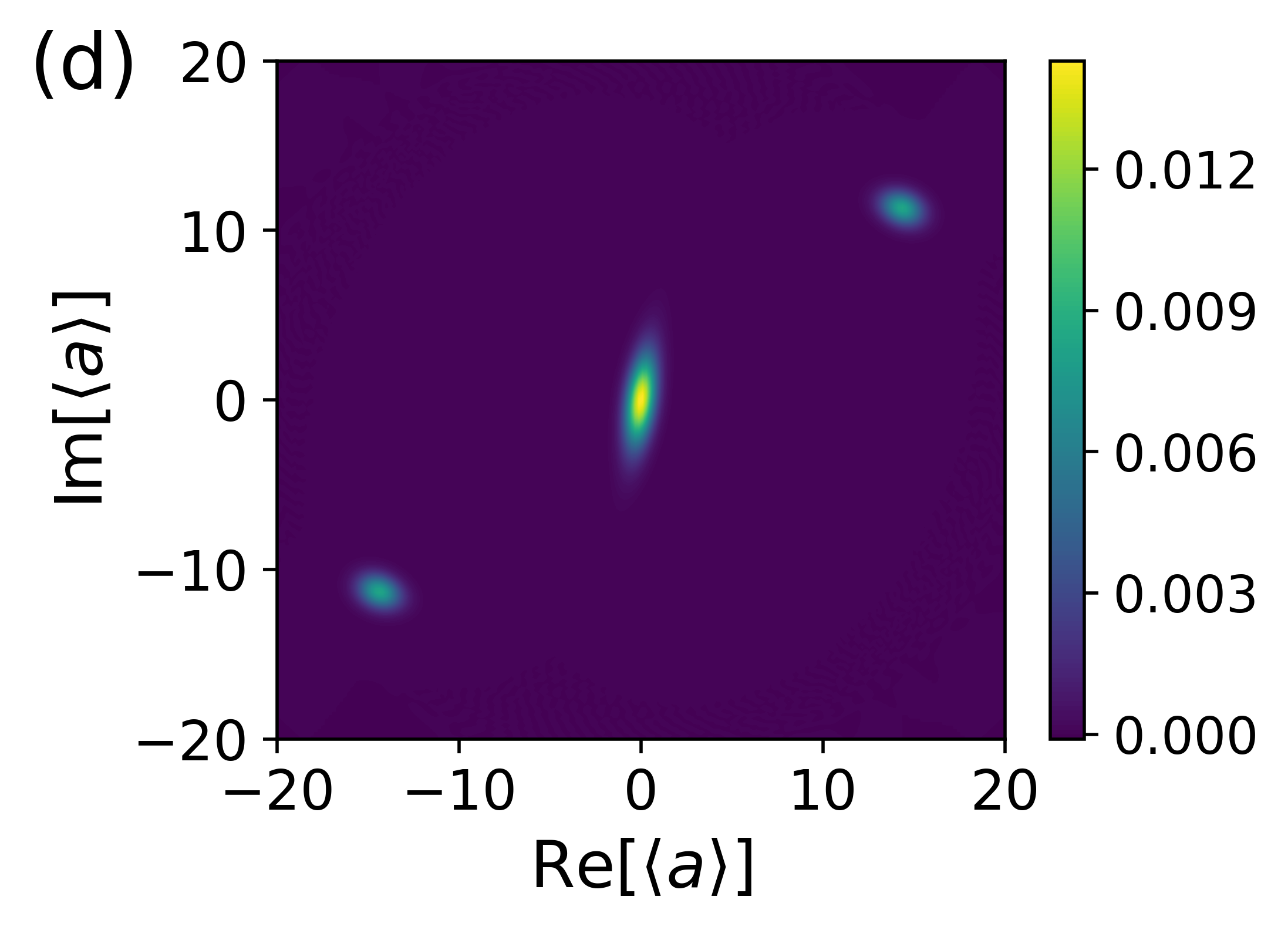}
	\caption{Wigner functions of the cavity field of the steady state for various $g_r$ and $g_{cr}$. (a) Normal phase. $g_r=0.43$ and $g_{cr}=0.7$; (b) Superradiant phase. $g_r=0.5$ and $g_{cr}=0.8$; (c, d) Bistable phase. $g_r=1.7$, $g_{cr}=0.55$ for (c) and $g_r=0.55$, $g_{cr}=1.7$ for (d). Here, we set $\eta =200$, the dimension of Fock space is $300$, and $\bar\kappa=0.5$.
	}
	\label{fig:WF}
\end{figure}

\begin{figure}[tpb!]
	\includegraphics[width=0.49\linewidth]{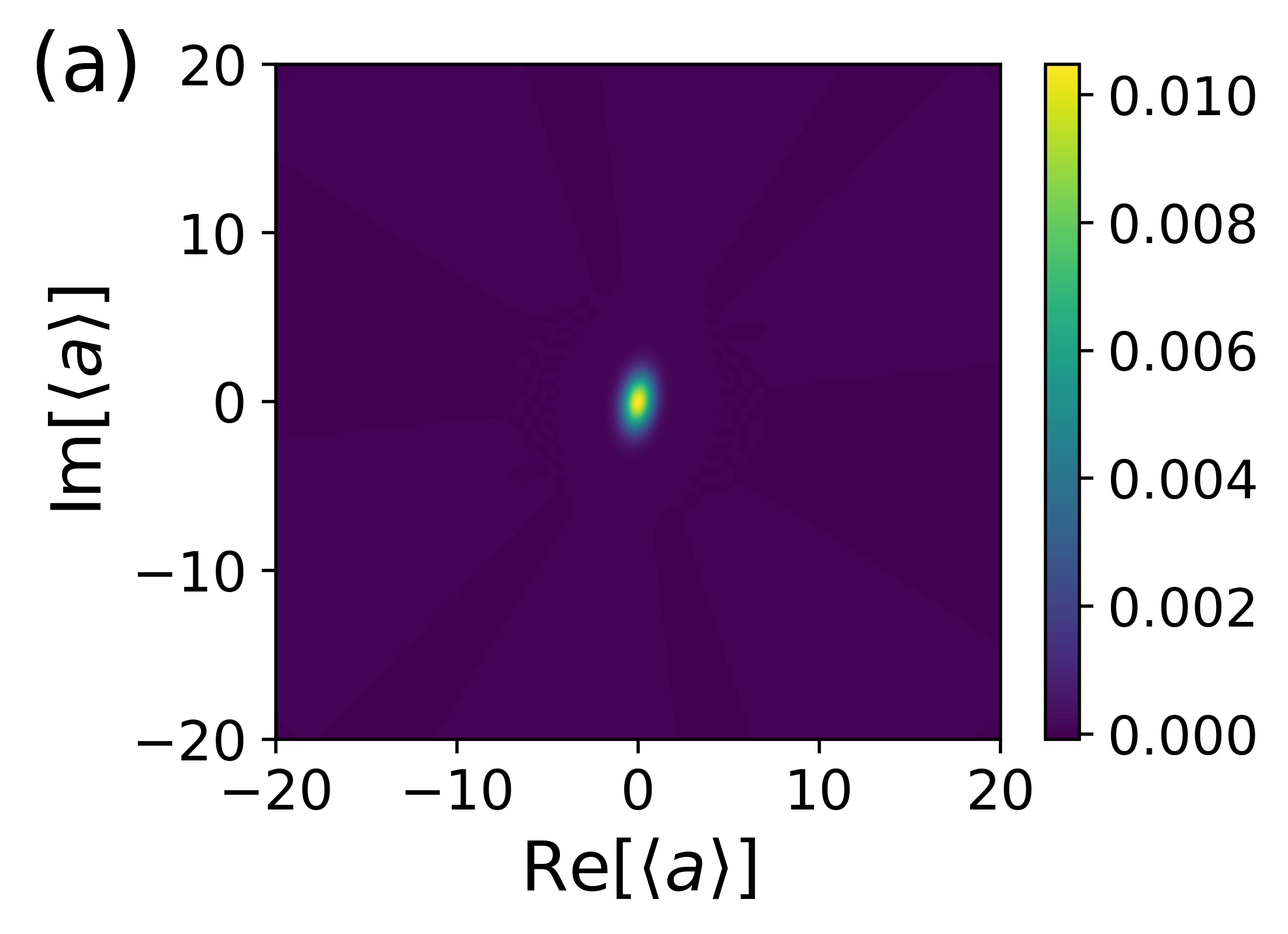}
	\includegraphics[width=0.49\linewidth]{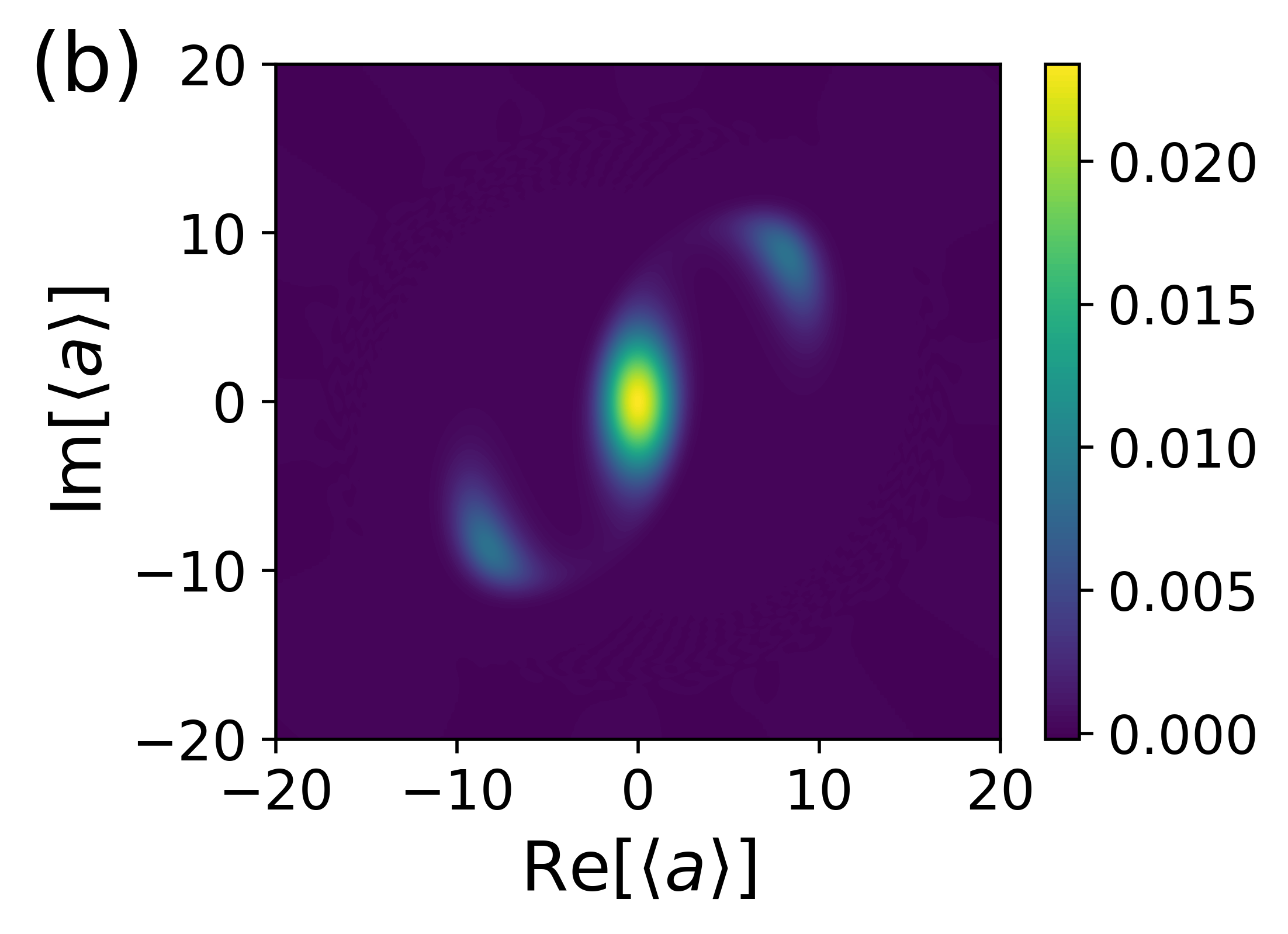}
	\caption{The role of spin damping on the phase boundaries. Wigner function of the photon part of the steady state for $ g_r=0.35 $ and $ g_{cr}=1.7$, corresponding to the NP shown in Fig.~\ref{fig:phasediagram}(c). (a) In the absence of the spin damping, the steady state is a Gaussian state centered at the origin. (b) In the presence of the spin damping whose strength is $\Gamma/\Omega=0.15$, the trimodal distribution appears indicating the bistability induced by the spin damping. Here, we set $\eta=200$, the dimension of Fock space is 300, and $\bar\kappa=0.5$.
	}
	\label{fig:WF2}
\end{figure}

In the main text, we have provided an exact solution for the master equation (\ref{eq:d_rho}) in the thermodynamic limit $\eta$$\rightarrow \infty$ using the mean-field analysis and the Schrieffer-Wolff transformation~\cite{Hwang2015PRL, Hwang2018PRA}. Away from the thermodynamic limit, there must be a unique steady-state solution. In this section, we show how the phase diagram featuring the bistable phase in the thermodynamic limit would manifest itself in this unique steady state using the numerical simulation of the quantum master equation (\ref{eq:d_rho}) for finite $\eta$ using QuTiP~\cite{QuTiP2012, QuTiP2013}. In Figs.~\ref{fig:WF} and~\ref{fig:WF2}, we show the Wigner function of the cavity field by projecting the steady state to the spin-down subspace.
In the NP, the cavity field is a squeezed state centered at the origin [see Fig.~\ref{fig:WF}(a)]. In the SP, the finite-$\eta$ induces tunneling between two superradiant solutions with opposite amplitudes, resulting in a superposition of two displaced squeezed states [see Fig.~\ref{fig:WF}(b)], referred to as a squeezed cat state. In the bistable phase, the tunneling between the normal and superradiant states for finite-$\eta$ leads to a trimodal steady state [see Figs.~\ref{fig:WF}(c) and \ref{fig:WF}(d)], which has also been observed in different open quantum systems showing bistability~\cite{Bartolo2016PRA, Landi2022PRA, Hines2005PRA,Stitely2020PRA}.

We note that the phase boundary of the NP and SP, and the resulting bistability phase is expected to be modified by the finite-frequency effects. Moreover, the trivial and stable solution with $s_z=+1$ can no longer be ignored when the spin frequency is finite. The fact that the $s_z=+1$ solution is stable for any $g$ and $\varepsilon$ is an artifact of introducing only the cavity damping to the master equation in Eq.~(\ref{eq:d_rho}), while neglecting spin damping. Therefore, when the finite spin frequency is considered, the spin damping should also be considered, which would further change the phase boundary. To demonstrate that the spin damping would modify the phase boundaries away from the thermodynamic limit, we compare a steady state in the absence and presence of the spin damping in Fig.~\ref{fig:WF2}. We choose $g_r=0.35$ and $g_{cr}=1.7$ where the steady state is a Gaussian state centered at the origin (NP) in the absence of the damping [see Fig.~\ref{fig:WF2}(a)]. When the spin damping is added, however, the steady state features a trimodal distribution which suggests the bistable phase [see Fig.~\ref{fig:WF2}(b)]. A more in-depth investigation into the role of spin damping on the phase diagram of the anisotropic open QRM is an interesting topic for future studies.

\section{Effective Hamiltonian for the Superradiant Phase} \label{App:Hsp_eff}
In this section, we derive an effective Hamiltonian for the SP. First, we apply a displacement unitary transformation to the original master equation~(\ref{eq:d_rho}), with the displacement operator $\mathcal{D}[\alpha_s]=\exp[\alpha_s a^\dagger -\alpha_s^* a]$. Setting that $\alpha_s$ equals the mean-field amplitude of the oscillator in the superradiant state, namely $\alpha_s=\alpha\equiv \braket{a}$, the master equation becomes
\begin{equation}
	\dot {\tilde{\rho}}=-i[\tilde H(\alpha),\tilde{\rho}]+\kappa(2a \tilde{\rho} a^\dagger - a^\dagger a\tilde{\rho}-\tilde{\rho} a^\dagger a),
\end{equation}
where $\tilde{\rho} \equiv \mathcal{D}^\dagger[\alpha_s] \rho\mathcal{D}[\alpha_s]$ and the effective unitary Hamiltonian $ \tilde H (\alpha) $ is given by
\begin{align}
	\tilde H (\alpha)&=\mathcal{D}^\dagger[\alpha]  H\mathcal{D}[\alpha]+i\kappa(\alpha^*a-\alpha a^\dagger)\nonumber \\
	& = H_s - V  + H_f ,
\end{align}
with the terms
\begin{align}
	H_s &  = \frac{\Omega}{2} \sigma_z - \sigma_+ (\lambda_r \alpha + \lambda_{cr}\alpha
	^*) - \sigma_- (\lambda_r \alpha^* + \lambda_{cr} \alpha),\\
	V & = \lambda_r (a\sigma_+ + a^\dagger \sigma_-) + \lambda_{cr}(a \sigma_- + a^\dagger \sigma_+),\\
	H_f & =\omega_0 \left[ a^\dagger a + a^\dagger \alpha (1 - i \bar{\kappa}) + a \alpha^* (1 + i \bar{\kappa}) + |\alpha|^2 \right].
\end{align}

%---------------------------------
%---------------------------------

%---------------------------------
%---------------------------------

We note that the spin part of the $H_s$ term can be diagonalized in terms of $U^\dagger H_s U = \frac{\Omega}{2} \sqrt{\xi}\sigma_z$ by the unitary operator
\begin{equation}
	\renewcommand{\arraystretch}{1.3}
	U = \frac{1}{\sqrt{2} \sqrt{\xi + \sqrt{\xi}}} \begin{pmatrix}
		1 + \sqrt{\xi} & 2(g_r \bar{\alpha} + g_{cr} \bar{\alpha}^* ) \\
		- 2(g_r \bar{\alpha}^* + g_{cr} \bar{\alpha})  & 1 + \sqrt{\xi}
	\end{pmatrix},
\end{equation}
where we have used the abbreviation $\xi \equiv 1 + 4|g_r \bar{\alpha}^* + g_{cr} \bar{\alpha}|^2=1/s_z^2$.
Correspondingly, the $V$ term becomes
\begin{align}
	U^\dagger V U  &=  (a \sigma_+ u^* + a^\dagger \sigma_- u) + (a \sigma_- v + a^\dagger \sigma_+ v^*) \nonumber\\
	&\quad - (a w + a^\dagger w^*) \sigma_z ,
\end{align}
where we have defined $u \equiv \sqrt{\Omega \omega_0} \bar{u}$, $v \equiv \sqrt{\Omega \omega_0} \bar{v}$ and $w \equiv \sqrt{\Omega \omega_0} \bar{w}$, and  $ \bar u $, $ \bar v $ and $ \bar w $ are given as follows:
\begin{align}
	\label{u_def}\bar u &= \frac{1}{2} \left( g_r \left(1 + \frac{1}{\sqrt{\xi}}\right) - g_{cr} \frac{(g_r \bar{\alpha}^* + g_{cr} \bar{\alpha})^2}{\xi + \sqrt{\xi}} \right),\\[0.5ex]
	\label{v_def}\bar v &= \frac{1}{2} \left( g_{cr} \left(1 + \frac{1}{\sqrt{\xi}}\right) - g_r \frac{(g_r \bar{\alpha}^* + g_{cr} \bar{\alpha})^2}{\xi + \sqrt{\xi}}\right), \\[0.5ex]
	\label{w_def}\bar w&= \frac{1}{\sqrt{\xi}} [g_r (g_r \bar{\alpha}^* + g_{cr} \bar{\alpha}) + g_{cr} (g_r \bar{\alpha} + g_{cr} \bar{\alpha}^*)].
\end{align}
Now, the effective unitary Hamiltonian $ \tilde H (\alpha) $ can be separated into diagonal and off-diagonal parts in the new spin basis
\begin{align}
	\tilde H(\alpha)&= H_d - H_{od},
	\label{eq:H_d_od}
\end{align}
with
\begin{align}
	H_d &= \omega_0 a^\dagger a + \frac{\Omega}{2} \sqrt{\xi} \sigma_z + \omega_0 |\alpha|^2+ H_l , \\
	H_{od} &= (a \sigma_+ u^* + a^\dagger \sigma_- u) + (a \sigma_- v + a^\dagger \sigma_+ v^*).
\end{align}
In the diagonal part, there is a term $ H_l $ which is linear in $a $ ($a^\dagger$):
$H_l = a^\dagger [\omega_0 \alpha (1-i\bar{\kappa}) + w^*\sigma_z] + a [\omega_0 \alpha^* (1+i \bar\kappa) + w \sigma_z]$.
We neglect this term in the following SW transformation, because it simply vanishes upon projection to the spin-down subspace.

Then, we perform the SW transformation to the Hamiltonian (\ref{eq:H_d_od}) with the generator
\begin{equation}
	\label{eq:S'}
	S' \simeq \frac{1}{\Omega\sqrt{\xi}}(u^* a \sigma_+ - u a^\dagger \sigma_-) + \frac{1}{\Omega\sqrt{\xi}} (v^* a^\dagger \sigma_+ - v a \sigma_-) ,
\end{equation}
where the approximation $1/(\Omega\sqrt{\xi} \pm \omega_0)  \simeq 1/\Omega\sqrt{\xi}$ is taken. By projecting it to the spin-down subspace, we finally obtain an diagonalized effective Hamiltonian for the SP,
\begin{align}
	\tilde H_\textrm{sp}= H_q + E_\text{sp},
\end{align}
where $E_\text{sp} = \omega_0 |\alpha|^2 - \frac{\Omega}{2  |s_z|}  - \omega_0 |s_z| |\bar v|^2$ is the ground energy,
and $ H_q $ has a quadratic from:
\begin{equation}
	H_q=P a^\dagger a +Q  a a  +Q^* a^\dagger a^\dagger,
\end{equation}
with the coefficients $ P$ and $ Q$ given as
\begin{align}
	\label{eq:P}P=& \omega_0 \left[1-|s_z|(|\bar{u}|^2 + |\bar{v}|^2)  \right],\\
	\label{eq:Q}Q=&-\omega_0|s_z| \bar{u}^* \bar{v}.
\end{align}

In addition, the expressions for $\bar u$ and $ \bar v $ given in Eqs.~(\ref{u_def}) and  (\ref{v_def}) can be further simplified by substituting the explicit expressions of $\bar{x}$ and $\bar{y}$ in Eqs.~(\ref{result_x}) and (\ref{result_y}):
\begin{align}
	\label{eq:u_ekg}
	\bar u &= \frac{g}{2} \left[ 1 + |s_z| - \frac{1}{2}(1 - |s_z|)\sqrt{4 \varepsilon^2 - \bar{\kappa}^2(1-\varepsilon^2)^2} \right] \nonumber\\
	&\quad+ i \frac{g}{4}(1 - |s_z|) |1-\varepsilon^2|  \bar{\kappa}, \\[0.5ex]
	\label{eq:v_ekg}
	\bar v &= \frac{g}{2} \left[ \varepsilon(1 + |s_z|) - \frac{1}{2\varepsilon}(1 - |s_z|)\sqrt{4 \varepsilon^2 - \bar{\kappa}^2(1-\varepsilon^2)^2} \right] \nonumber\\
	&\quad+ i \frac{g}{4\varepsilon}(1 - |s_z|) |1-\varepsilon^2|  \bar{\kappa}.
\end{align}

\bibliography{Multicritical_DPT_refs}

\end{document}